\documentclass[11pt]{article}

\usepackage[T1]{fontenc}

\usepackage[nottoc,numbib]{tocbibind}
\usepackage[margin=1.2in]{geometry}
\usepackage{float}
\usepackage[english]{babel}
\usepackage[utf8]{inputenc}
\usepackage{amsmath}
\usepackage{listings}
\usepackage{graphicx}
\usepackage[colorinlistoftodos]{todonotes}

\begin{document}

\newcommand*{\getTitleGer}{Intrinsic Instabilities and Mechanical Anisotropy in Halide Perovskite Monolayers}
\newcommand*{\getFaculty}{\textit{$^{a}$~Departamento de Física dos Materiais e Mecânica, Instituto de Física, Universidade de São Paulo, São Paulo 05508-090, São Paulo, Brazil.} \newline
\textit{$^{b}$~INFIQC, CONICET, Departamento de Química Teórica y Computacional, Facultad de Ciencias Químicas, Universidad Nacional de Córdoba, Argentina.}
\newline
\textit{$^{c}$~Departamento de Física dos Materiais e Mecânica, Instituto de Física, Universidade de São Paulo, São Paulo 05508-090, São Paulo, Brazil.}}
\newcommand*{\getAuthor}{Gabriel X. Pereira,\textit{$^{a}$} Lucas M. Farigliano,\textit{$^{a}$}\textit{$^{b}$} Roberto H. Miwa,\textit{$^{c}$}, and Gustavo M. Dalpian\textit{$^{\ast a}$}}
\newcommand*{\getContact}{Contact}

\begin{titlepage}
    \oddsidemargin=\evensidemargin\relax
    \textwidth=\dimexpr\paperwidth-2\evensidemargin-2in\relax
    \hsize=\textwidth\relax
    
    \centering

    \line(1,0){400}

    \vspace{0,5cm}
    
    {\huge \getTitleGer{}}

    \vspace{1cm}
    
    \getAuthor{}
    \vspace{0,25cm}

    \getFaculty{}
    \vspace{0,25cm}

    \begin{abstract}
    Halide perovskites have been extensively studied owing to their excellent optoelectronic properties and their unique lattice characteristics, that are very soft and anharmonic. Recent studies indicate the importance of a deep understanding of their surfaces and, in the limit, the properties of low-dimensional structures based on these materials. To investigate the structural and electronic properties of halide perovskite monolayers (i.e., perovskenes), this work uses first-principles simulations. We have studied three different stoichiometries (ABX$_3$, ABX$_4$, and A$_2$BX$_4$) and structural phases for iodide, bromide, and chloride perovskite monolayers. Their thermodynamic behavior was evaluated through the construction of phase diagrams, highlighting the instability of the ABX$_4$ stoichiometry, which was further supported by its mechanical instability. Structurally, the covalent characteristics of the Pb--X bond, in contrast to the Cs--X bonds, induce a strong anisotropy in the Young’s modulus and Poisson’s ratio along different crystallographic directions, and also account for the lower stiffness observed in the phases where the octahedra are not aligned. The electronic properties are somewhat similar to those of their 3D counterparts, but with a slightly larger band gap; in the monolayers, the band gap increases with halogen electronegativity (I, Br, Cl) and octahedral tilting. Moreover, the non-symmetric ABX$_3$ stoichiometry exhibited a spin splitting due to the internal dipole moment in these layers. Overall, our work lays the groundwork for a deeper understanding of low-dimensional structures based on halide perovskites.
    \end{abstract}
  
\end{titlepage}

\section{Introduction}

As a widely studied class of materials, halide perovskites have attracted great interest for the fabrication of optoelectronic devices \cite{dong_ran_gao_li_xia_huang_2023}. Their electronic properties, such as high carrier mobility and long carrier diffusion lengths, combined with high absorption coefficients and the possibility of band-gap tuning through chemical and/or structural modifications \cite{manser_christians_kamat_2016}, enhance their potential applications in photovoltaic cells, light-emitting diodes (LEDs), photocatalysts, and photodetectors \cite{jena_2019, liu_ma_zhang_he_dai_li_shi_manna_2025, huang_yu_wu_li_li_2024, morais_caturello_lemes_ferreira_ferreira_acuña_brochsztain_dalpian_souza_2024, zheng_pauporté_2023}. Emerging phenomena, such as ferroelectricity and topological phases \cite{shamim_2020, zheng_2022, jin_im_freeman_2012, lafuente-bartolome_lian_giustino_2024}, also expand the applicability of these materials into less conventional application areas \cite{kim_han_choi_kim_jang_2018}.

Structurally, halide perovskites are defined by a crystal lattice composed of a network of corner-sharing $BX_6$ octahedra, while the A-site atom is usually enclosed within an $AX_{12}$ dodecahedron \cite{akkerman_manna_2020}. In addition, it is common for the octahedral framework to form a tilting pattern \cite{wang_wang_doherty_stranks_gao_yang_2025, zhao_dalpian_wang_zunger_2020}, as this behavior is strongly related to structural stability, especially due to the A-cation interactions, with the degree of tilting being related to the size of the A-site atom (Cs, Sn, MA, FA) \cite{hautzinger_mihalyi-koch_jin_2024}. 

In low-dimensional systems, halide perovskites can exhibit novel or enhanced properties that strongly depend on their size and morphology \cite{chu_chu_zhao_ye_jiang_zhang_you_2021, Leite2026Oct}. For optoelectronic applications, low-dimensional halide perovskites stand out due to their low trap-state density, tunable band gap, and high photoluminescence quantum yield \cite{kim_park_cho_jang_2024, rafique_abbas_mendes_barquinha_martins_fortunato_águas_jana_2025}. In two dimensions, halide perovskites can be divided into quasi-2D perovskites and few-layer perovskites \cite{laxmi_2025}, also referred to as perovskenes \cite{D3CP04435A, FARHADI2025113581}. 

Quasi-2D perovskites are characterized by a hybrid structure composed of an inorganic layer, with stoichiometry $A_{n-1}B_nX_{3n+1}$, and an organic layer that typically binds to the A-site of the inorganic framework \cite{mao_stoumpos_kanatzidis_2018, wu_liang_zhang_ge_xing_sun_2021}. The organic component in quasi-2D perovskites usually does not significantly affect the electronic structure near the Fermi level, so that some of the electronic properties typical of truly two-dimensional materials are also observed in quasi-2D perovskites, despite their overall three-dimensional nature \cite{milot_2016}. Conversely, perovskenes correspond to cases where these materials are atomically thin along one crystallographic direction and can exist in the stoichiometries ABX$_3$, ABX$_4$, and A$_2$BX$_4$ \cite{ricciardulli_yang_smet_saliba_2021}, presented in the Figure~\ref{fig:fig1}.(a-c). This feature makes perovskenes distinct from their bulk inorganic counterparts not only in terms of their electronic structure, but also in their morphology and structural properties \cite{thakur_chang_2023}.

In recent years, several synthesis routes have been proposed for the preparation of halide perovskite thin films, with particular emphasis on hybrid perovskites \cite{orange_tambwe_arendse_malevu_ross_2024, li_zhu_wang_pan_cao_wu_chen_2023}, along with important advances also in all-inorganic two-dimensional halide perovskites, especially in the synthesis of nanoplatelets \cite{yehonadav_2015, akkerman_motti_2016}. The top-down approaches mainly rely on mechanical exfoliation of bulk layered perovskite crystals \cite{niu_2014, yaffe_2015}, whereas the bottom-up routes involve liquid-phase and epitaxial growth techniques \cite{weidman_seitz_stranks_tisdale_2016, hu_2024}. The continuous development of these methodologies has enabled the fabrication of increasingly thinner films, reaching the atomic limit \cite{dou_2015, leng_2018, shi_2020}.

In parallel with experimental developments, several recent computational studies have sought to explore and understand these materials. Regarding the discovery of new perovskite monolayers, J. Yang \textit{et al.} proposed a high-throughput screening to evaluate different compositions and thicknesses of inorganic 2D perovskites in the cubic phase, without accounting for octahedral tilting \cite{yang_li_2022}. Conversely, C. Ming \textit{et al.} reported computational results for two-dimensional Ba--Zr--S systems, as well as comparisons with an iodide perovskite monolayer of A$_2$BX$_4$ stoichiometry, highlighting the importance of octahedral tilting for the electronic properties of these materials \cite{ming_yang_zeng_zhang_sun_2020}. Meanwhile, J. Zhang \textit{et al.} investigated the influence of octahedral disorder on the ferroelectricity of Sn- and Pb-based halide perovskenes with A$_2$BX$_4$ stoichiometry \cite{zhang_xie_he_ji_shen_2024, zhang_ji_he_xie_2024}.

Building upon these recent theoretical and experimental advances, the present work aims to evaluate the stability of different compositions, stoichiometries, and phases of halide perovskenes, as well as their mechanical and electronic properties. Accordingly, the thermodynamic stability of the ABX$_3$, ABX$_4$, and A$_2$BX$_4$ stoichiometries was compared \cite{ricciardulli_yang_smet_saliba_2021}, together with the structural stability of distinct crystalline phases characterized by different PbX$_6$ octahedral tilting patterns. Furthermore, the mechanical and thermodynamic stabilities were analyzed as a function of the halogen species (I, Br, or Cl) in the composition. Regarding the mechanical response, several elastic properties were investigated, with particular attention to the role of octahedral tilting. Finally, the electronic structure characteristics were examined both as a means to interpret the structural and mechanical behaviors and to assess potential applications and emerging functionalities in these materials.

\section{Results and Discussion}

\begin{figure*}[h!]
    \centering
    \includegraphics[width=1\linewidth]{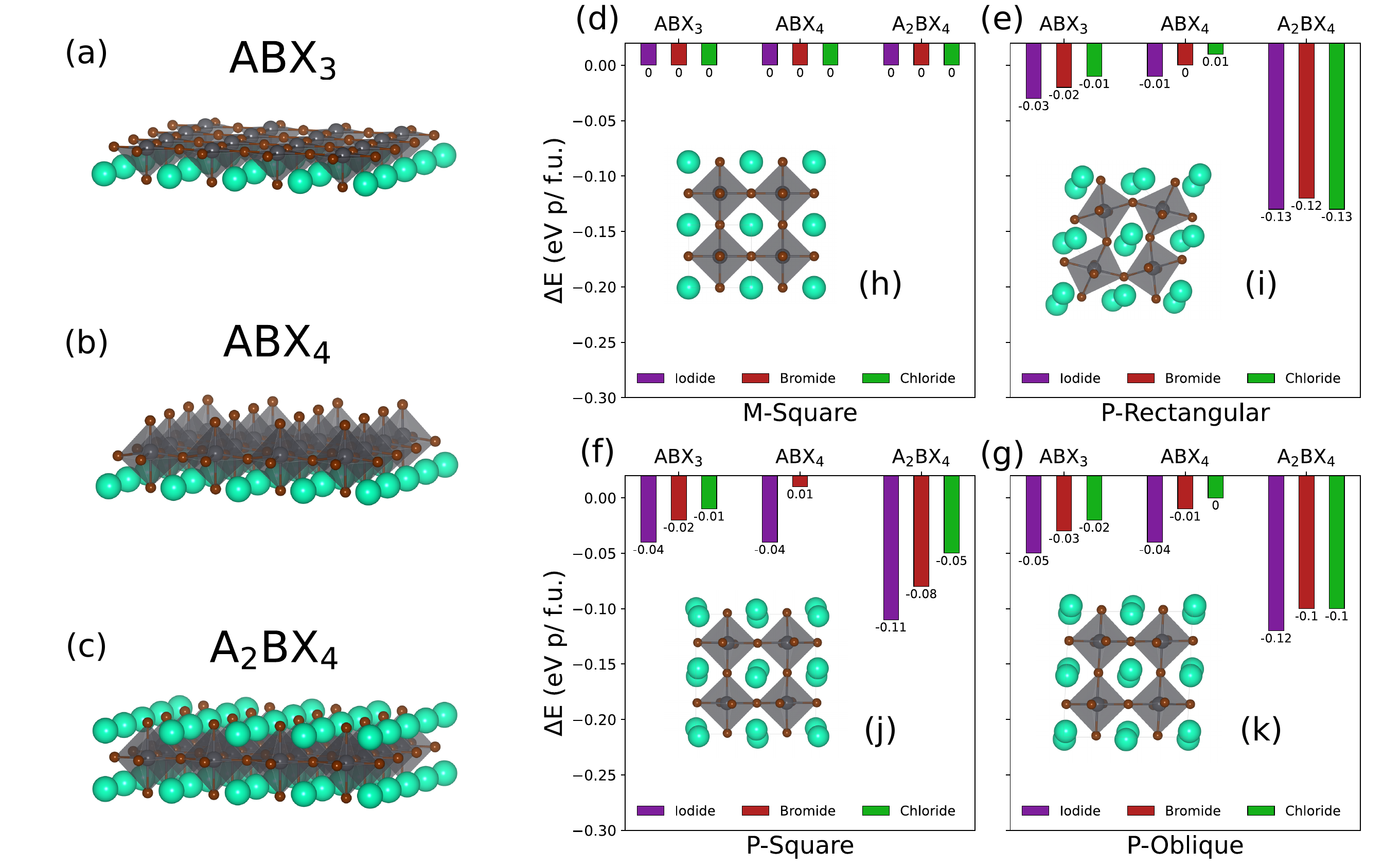}
    \caption{Structural difference between the (a) ABX$_3$, (b) ABX$_4$, and (c) A$_2$BX$_4$ monolayers. Energy differences relative to the M-Quadratic phase for the phases (d) M-Quadratic, (e) P-Rectangular, (f) P-Square, and (g) P-OBlique. The thermodynamic stability of the octahedral tilting is evidenced. Each color represents a halide, as showed in the labels. Finally, each stoichiometry is grouped in sets of 3 bars. The most stable phases for ABX$_3$ and ABX$_4$ are the P-Oblique structures, whereas for A$_2$BX$_4$ the P-Rectangular phase is favored. Atomic structures of each studied phase are shown in panels (e--h).}
    \label{fig:fig1}
\end{figure*}

To evaluate the stability of halide perovskite monolayers, it is first necessary to establish how these materials are structurally organized. In this sense, in line with the behavior extensively described for three-dimensional (3D) halide perovskites \cite{lee_bristowe_lee_lee_bristowe_cheetham_jang_2016}, the results presented in Figure~\ref{fig:fig1} show that the monolayers also exhibit, as a characteristic of their stable structures, a tilting pattern of PbX$_6$ octahedra. Four different crystal structures were analyzed: (i) The M-Square structure represents the $1\times1$ cell, which does not allow octahedral tilting, calculated considering the cell size optimization. (ii) The P-Rectangular represents the structure obtained from a $\sqrt{2} \times \sqrt{2}$ cell that allows octahedral tilting, and by considering the optimization of the lattice vectors during the simulation. (iii) The P-Square is a structure obtained from a $2 \times 2$ cell, allowing just the optimization of the cell size without changing the cell symmetry. (iv) The P-Oblique represents a structure calculated from a $2 \times 2$ cell, allowing both the optimization of cell size and symmetry. It generally resulted in oblique cells, so that the obliquity depended on the stoichiometry and composition of each layer. For instance, the angle between the in-plane vectors for Cs$_2$PbCl$_4$ is $\approx 83.45^\circ$, while for CsPbBr$_3$ is $\approx 89.88 ^\circ$. The M- refers to the monomorphic phases, while P- corresponds to the polymorphic phases.

The energetic analysis further shows that, although a tilting pattern is the most stable configuration for all halide perovskite monolayers studied, each stoichiometry displays distinct behaviors regarding long-range ordering. While ABX$_3$ monolayers tend to slightly break the square symmetry of the unit cell (P-Oblique, Fig.~\ref{fig:fig1}g), A$_2$BX$_4$ monolayers present low-energy regions where the oblique tendency is much more pronounced (P-Rectangular, Fig.~\ref{fig:fig1}h). On the other hand, the ABX$_4$ stoichiometry, despite the instability discussed in the following sections, also exhibits the oblique structure as its most stable configuration.

Moreover, concerning the possible crystalline phases of perovskite monolayers, it is important to note that the energy differences between the phases are on the order of 10~meV per formula unit, indicating that phases with long-range symmetry but local symmetry breaking could be thermally accessible, as previously detailed for 3D halide perovskites \cite{farigliano_ribeiro_dalpian_2024, bonadio_2021}. To further confirm this behavior, ab initio molecular dynamics (AIMD) simulations were performed at 100, 200, and 300 K for all three stoichiometries considered in this work (ABX$_3$, ABX$_4$, and A$_2$BX$_4$), and for both quadratic and oblique phases, with a $4 \times 4$ superc
ell. All simulations were carried out within the canonical NVT ensemble, keeping the cell volume fixed at the value obtained from the fully relaxed structures. Each trajectory had a total duration of 15 ps. After completing the AIMD runs, additional structural relaxations were performed in order to assess whether the systems could recover the configurations analyzed in the rest of this study.

Our results indicate that the ABX$_3$ and A$_2$BX$_4$ systems are thermally stable over the entire simulation time. In particular, for the oblique cells, the structures recovered after relaxation closely resemble the configurations obtained from the initial energy minimization. In contrast, for the quadratic cells, the relaxed structures exhibit local distortions while preserving the overall perovskite framework. This behavior is consistent with what has been previously reported for three-dimensional perovskite materials \cite{zhao_dalpian_wang_zunger_2020}. On the other hand, the ABX$_4$ systems show poor thermal stability at finite temperatures. Significant structural distortions are observed during the simulations, involving changes in the bonding environment of halogen atoms that are not coordinated to nearby Cs atoms, ultimately preventing the recovery of the original perovskite-like structure.

\begin{figure*}[h]
    \centering
    \includegraphics[width=1\linewidth]{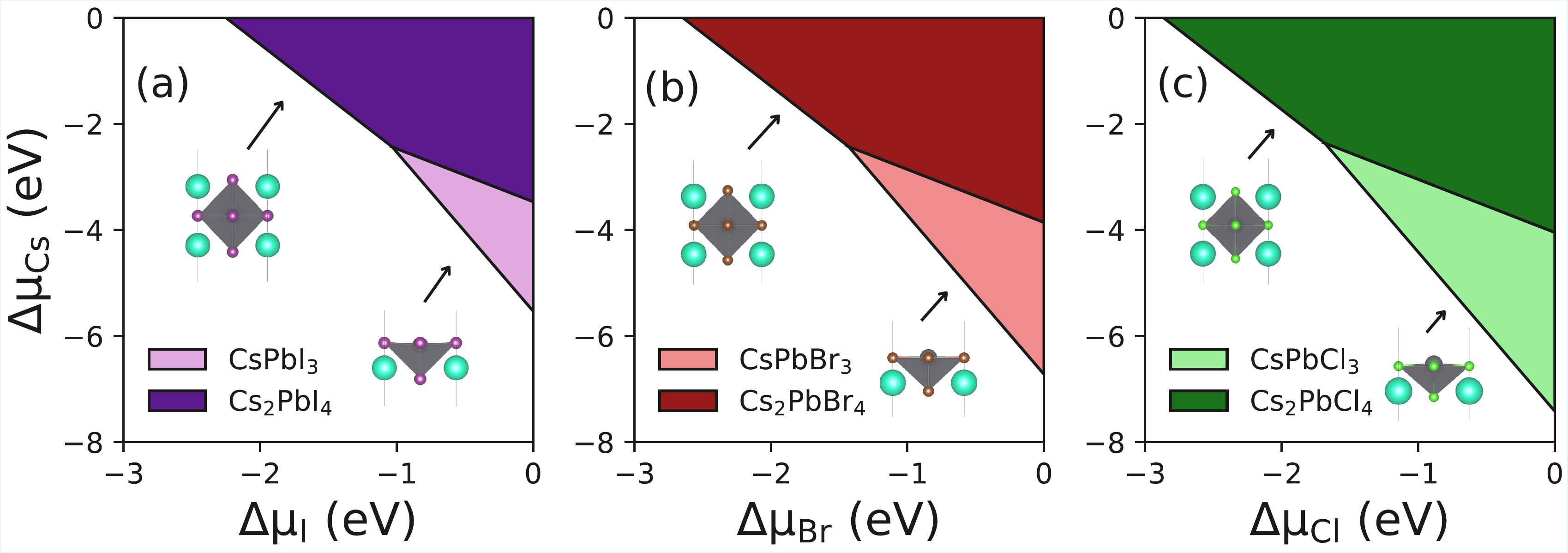}
    \caption{Phase diagrams comparing stoichiometries. Each diagram represents the behavior of (a) Iodides, (b) Bromides, and (c) Chlorides. The filled areas represent the regions where each stoichiometry is thermodynamically stable. These results indicate the thermodynamic instability of the ABX$_4$ stoichiometry.}
    \label{fig:fig2}
\end{figure*}

Complementary to the analysis of phase stability, the thermodynamic stability with respect to stoichiometry and elemental chemical potentials was also evaluated. As an alternative approach to comparing the different stoichiometries, variations in the chemical potentials of the halogens and cesium were used as parameters to determine stability under Cs- or X-rich ($\Delta \mu_{(Cs, X)} = 0$) and Cs- or X-poor conditions (X = I, Br, or Cl). Accordingly, the results presented in Figure~\ref{fig:fig2} not only reveal the thermodynamic stability of the A$_2$BX$_4$ stoichiometry under Cs- and X-rich conditions, which is expected due to the higher content of these elements, but also predict an A$_2$BX$_4$\,→\,ABX$_3$ stoichiometric transition upon the reduction of the Cs chemical potential under halogen-rich conditions.

Moreover, it is important to emphasize that the thermodynamic instability of the ABX$_4$ stoichiometry does not preclude the possibility of isolated Cs vacancies within the monolayer structure, but rather reinforces the instability of halogens when they lose coordination with cesium atoms. On other hand, as described in the Introduction, halide perovskite monolayers or thin films prepared experimentally are constantly subjected to external stresses, either from the substrate or from heterojunctions in microdevices. Therefore, to extend the study toward the applicability of these materials, it is necessary to understand not only their stability but also their behavior under different mechanical stresses.

In this context, the first noteworthy mechanical feature of halide perovskite monolayers is their low stiffness when compared to oxide perovskite monolayers \cite{D3CP04435A}. While M.~Naseri \textit{et al.} computationally showed that SrZrO$_3$, BaTiO$_3$, and BaZrO$_3$ monolayers exhibit Young’s moduli of respectively 125.54–154.60~$\mathrm{N}\cdot\mathrm{m}^{-1}$, 144.03–152.39~$\mathrm{N}\cdot\mathrm{m}^{-1}$, and 104.32–170.65~$\mathrm{N}\cdot\mathrm{m}^{-1}$, the results presented in Table~\ref{tab:mechprops} demonstrate that halide perovskites have Young’s moduli approximately one order of magnitude smaller, which is in agreement with the well-known soft nature of halide perovskites \cite{Ferreira2018Aug}. Moreover, in general, monolayers with tilted octahedra display Young’s moduli nearly twice as small as those of monomorphic monolayers, highlighting the influence of octahedral dynamics on the stiffness of these materials.  

Beyond their effect on stiffness, octahedral tilting also considerably impacts the Poisson’s ratio of the monolayers, particularly along the Pb–X bonding direction. As shown in Table~\ref{tab:mechprops}, for monomorphic phases, without octahedral tiltings, these materials exhibit very small Poisson’s ratios, which can even become negative in the case of iodide perovskites. For polymorphic phases, that presents octahedral tiltings, however, a noticeable increase in the Poisson’s ratio is observed, although values along the Pb–X bonds remain close to zero for all halogens and stoichiometries. 

\begin{table*}
    \centering
    \setlength{\tabcolsep}{1pt}
     \caption{Elastic properties of halide perovskite monolayers. The Young's modulus (Y$_{2D}$), Shear modulus (G$^{2D}$), and Layer modulus (Lm) are in $N \cdot \mathrm{m}^{-1}$. The Poisson’s ratio ($\nu$) is dimensionless. The last two rows indicate if the monolayers are mechanically stable, according to the elastic stability conditions for 2D materials. The M- indicates the M-Square phase and P- refers to the stablest between the P-Rectangular, P-Square and P-Oblique phases, being that the rest are presented in the SI.}
    \begin{tabular}{cccccccccc}
    \hline
    Property \;\; & CsPbI$_3$ & CsPbI$_4$ & Cs$_2$PbI$_4$ & CsPbBr$_3$ & CsPbBr$_4$ & Cs$_2$PbBr$_4$ & CsPbCl$_3$ & CsPbCl$_4$ & Cs$_2$PbCl$_4$ \\
    \hline
    M-Y$^{2D}$ & 24.45 & 28.94 & 29.87 & 11.94 & 21.77 & 27.38 & 34.50 & 39.03 & 42.98 \\
    P-Y$^{2D}$ & 16.99 & -- & 19.51 & 17.50 & 9.88 & 21.92 & 21.58 & 27.05 & 22.87 \\
    M-$\nu^{2D}$ & -0.04 & 0.00 & 0.01 & -0.13 & 0.05 & 0.08 & 0.01 & 0.02 & 0.03 \\
    P-$\nu^{2D}$ & 0.14 & -- & 0.07 & 0.37 & 0.40 & 0.09 & 0.28 & 0.07 & 0.05 \\
    M-G$^{2D}$ & 2.13 & 2.33 & 2.64 & 2.96 & 3.02 & 3.53 & 1.26 & 1.29 & 1.42 \\
    P-G$^{2D}$ & 2.02 & -- & 9.12 & 2.17 & 10.06 & 0.72 & 2.22 & -3.15 & 10.91 \\
    M-Lm & 11.71 & 14.49 & 15.16 & 5.29 & 11.41 & 14.86 & 17.43 & 19.84 & 22.20 \\
    P-Lm & 11.70 & -- & 10.48 & 13.96 & 12.04 & 8.59 & 14.94 & 14.59 & 12.01 \\
    M-Stable & True & True & True & True & True & True & True & True & True \\
    P-Stable & True & False & True & True & True & True & True & False & True \\
    \end{tabular}
    \label{tab:mechprops}
\end{table*}

Regarding the shear modulus, as shown in Table~\ref{tab:mechprops}, for the ABX$_3$ stoichiometry—which deviates only slightly from square symmetry in the P-Oblique (most stable) phase—the $G_\text{2D}$ value decreases slightly compared to the monomorphic phase. For the other stoichiometries, a more significant variation of this modulus is found, which can be attributed to its relation to the oblique–square structural transition. Similarly to the shear modulus, the layer modulus also exhibits a marginal decrease in ABX$_3$ monolayers and a more pronounced decrease for the ABX$_4$ and A$_2$BX$_4$ stoichiometries. Beyond the mechanical properties of the perovskene monolayers, the analysis of the stress–strain tensor also allows the assessment of their mechanical stability \cite{born1988dynamical}. In this context, Table \ref{tab:mechprops} also reports the stability information for each monolayer investigated.

\begin{figure*}[h!]
    \centering
    \includegraphics[width=1\linewidth]{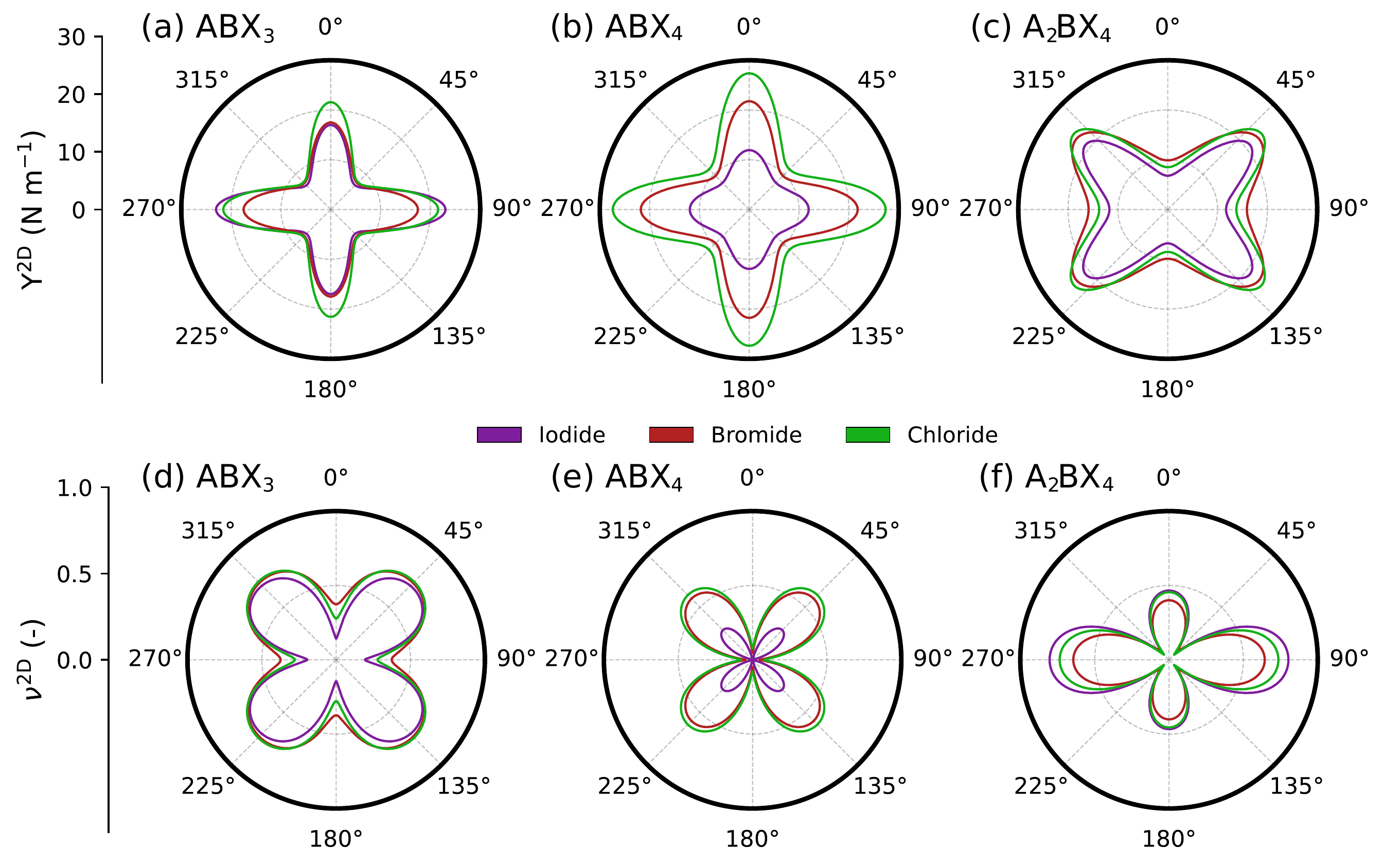}
    \caption{Young’s modulus ($Y^{2D}$) and Poisson’s ratio ($\nu^{2D}$) for (a, d) ABX$_3$ (P-Oblique), (b, e) ABX$_4$ (M-Quadratic), and (c, f) A$_2$BX$_4$ (P-Rectangular), respectively. The angles represent the different crystalline directions and the colors represent the different halogens, which do not significantly affect the mechanical properties. Octahedral tilting and monolayer stoichiometry play a crucial role in determining both Young’s modulus and Poisson’s ratio. In the Square and Oblique phases, the Pb-X bonds are aligned along the 0$^\circ$, 90$^\circ$, 180$^\circ$, and 270$^\circ$ directions, and the X-X bonds are in the 45$^\circ$, 135$^\circ$, 225$^\circ$, and 315$^\circ$ directions. In the rectangular phases, the behavior is inverted, with the X-X bonds along the 0$^\circ$, 90$^\circ$, 180$^\circ$, and 270$^\circ$, and the Pb-X bonds in the 45$^\circ$, 135$^\circ$, 225$^\circ$, and 315$^\circ$ directions.}
    \label{fig:fig3}
\end{figure*}

In addition to the behavior of mechanical properties with respect to each stoichiometry or halogen, it is also important to analyze how these properties vary according to the crystallographic direction. In this regard, the plots presented in Figure~\ref{fig:fig3}(a–c) show the magnitude of the Young’s modulus along each crystallographic direction of the monolayer. The first notable feature is that the perovskites are considerably stiffer (softer) along the Pb–X (X–X) bond direction, here characterized by maximum (minimum) values of $Y^\text{2D}$.

On the other hand, regarding the Poisson’s ratio (Fig.~\ref{fig:fig3}(d–f)), it is evident that along the Pb–X bond direction the value is significantly lower than along the X–X bond direction. In practice, this behavior indicates that when the Pb–X bonds are deformed along one direction, the Pb–X bonds in the orthogonal direction are barely affected. Conversely, when deformation occurs along the X–X bonds, a large deformation arises in the orthogonal X–X direction, altering the X–Pb–X angles (bending of the Pb-X bonds) and thus partially preserving the Pb–X bond lengths (stretching of the Pb-X bonds).

This set of mechanical characteristics can be understood as a direct consequence of the fact that, in the structure of halide perovskites, the Pb–halogen bonds are more rigid than the other interactions \cite{sun_isikgor_deng_wei_kieslich_bristowe_ouyang_cheetham_2017}. This rigidity mainly arises from the overlap between the Pb $s$ and $p$ orbitals and the halide $p$ orbitals, which imparts a covalent character—typically stiffer—to the Pb–X bond, despite the overall ionic nature of the material \cite{walsh_2015, min_2023}.

Complementarily, the higher stiffness of the Pb–X bonds also partially explains the tilting dynamics of the PbX$_6$ octahedra. The lower stiffness of the polymorphic phases is largely due to the fact that the octahedra can rearrange in a way that accommodates the total deformation without significantly distorting the Pb–X bonds, thereby reducing the total elastic energy and, consequently, the Young’s modulus.  

From an application perspective, the softness associated with tilting dynamics can facilitate the adaptation of perovskite monolayers to different substrates and the epitaxial growth on solid surfaces, as deformations in these materials do not result in large stresses \cite{jiang_zheng_liu_wang_2019}. Moreover, the mechanical compliance of these materials may also be a favorable property for the formation of heterobilayers or twisted bilayers \cite{chen_liu_li_cheng_ma_wang_li_2020, zhang_2024, Araujo2025Dec}.

\begin{figure*}[h!]
    \centering
    \includegraphics[width=1\linewidth]{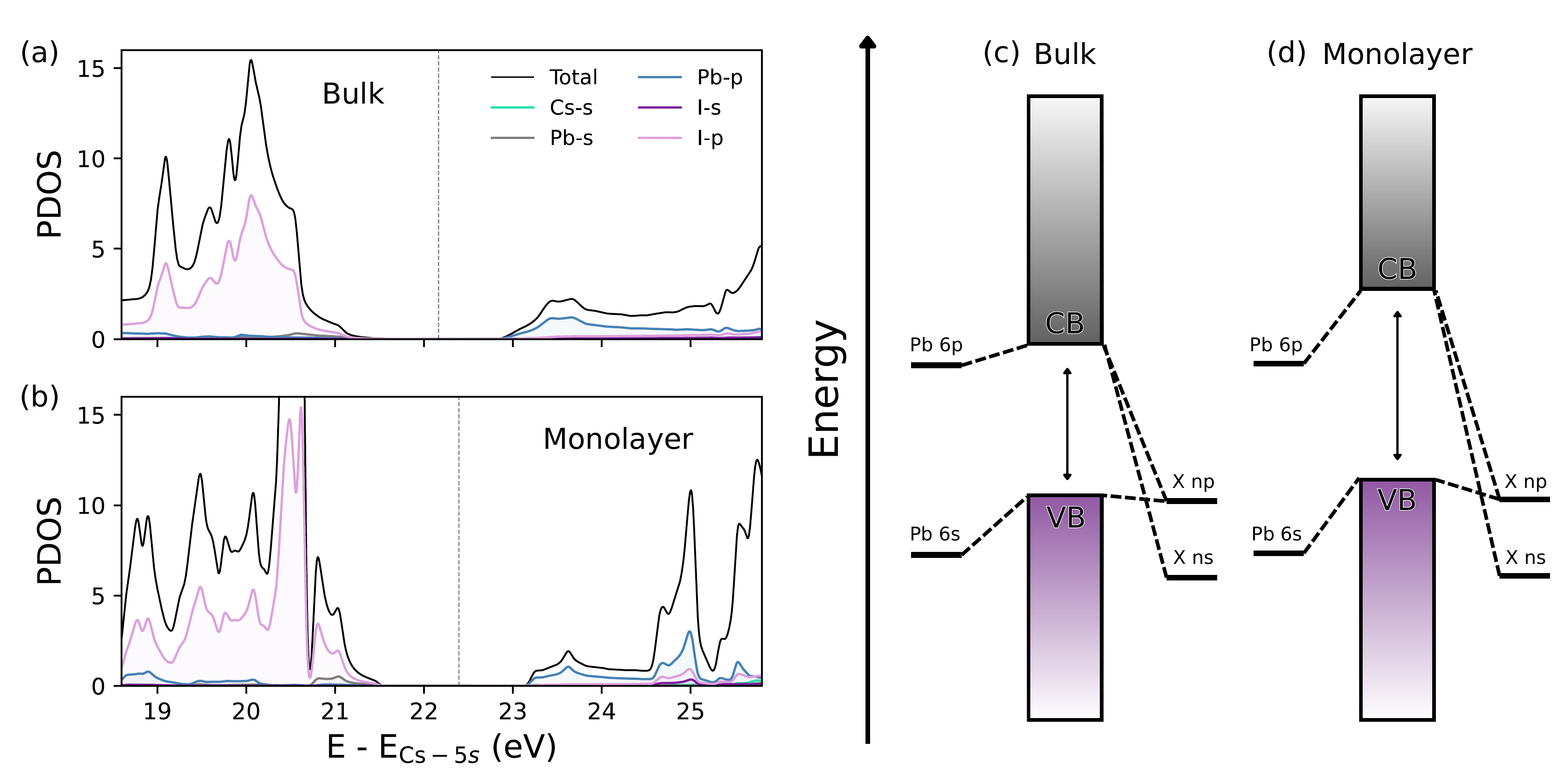}
    \caption{Partial density of states for CsPbI$_3$ bulk (a) and Cs$_2$PbI$_4$ monolayer (b) at PBE functional level. Schematic representation of molecular orbitals for (c) bulk and (d) monolayer halide perovskites. The states were aligned by the energy of the Cs-5s orbital.}
    \label{fig:fig4}
\end{figure*}

Finally, to further investigate the monolayers of these materials and explore their potential applications, it is necessary to evaluate the characteristics of the electronic structure of the perovskenes. In this regard, a comparison between the bulk and monolayer forms provides important insight into their electronic properties. Figures~\ref{fig:fig4}(a) and ~\ref{fig:fig4}(b) presents the partial density of states (PDOS) for bulk CsPbI$_3$ and the Cs$_2$PbI$_4$ monolayer. In both cases, the conduction band is primarily composed of Pb $6p$ and I $5p$ orbitals, with minor contributions from I $5s$ states, whereas the valence band is mainly formed by I $5p$ and Pb $6s$ orbitals, with a small Pb $6p$ contribution. This hybridization between Pb $s$ and $p$ orbitals and halogen $p$ orbitals supports the interpretation that the Pb--X bond in perovskenes possesses a partial covalent character, which underlies several of the mechanical and structural properties discussed previously \cite{min_2023}.

Despite the overall similarity in orbital composition, the bulk and monolayer systems differ slightly in both band gap magnitude (by approximately $0.2$~eV) and band alignment. These differences can be attributed to a combination of quantum confinement effects and the presence of dangling Pb--X bonds at the monolayer surfaces \cite{tang_2021}. Figures~\ref{fig:fig4}(c) and~\ref{fig:fig4}(d) schematically illustrate the orbital contributions to the valence and conduction bands and the associated increase in the band gap upon dimensional reduction.

\begin{figure*}[h!]
    \centering
    \includegraphics[width=1\linewidth]{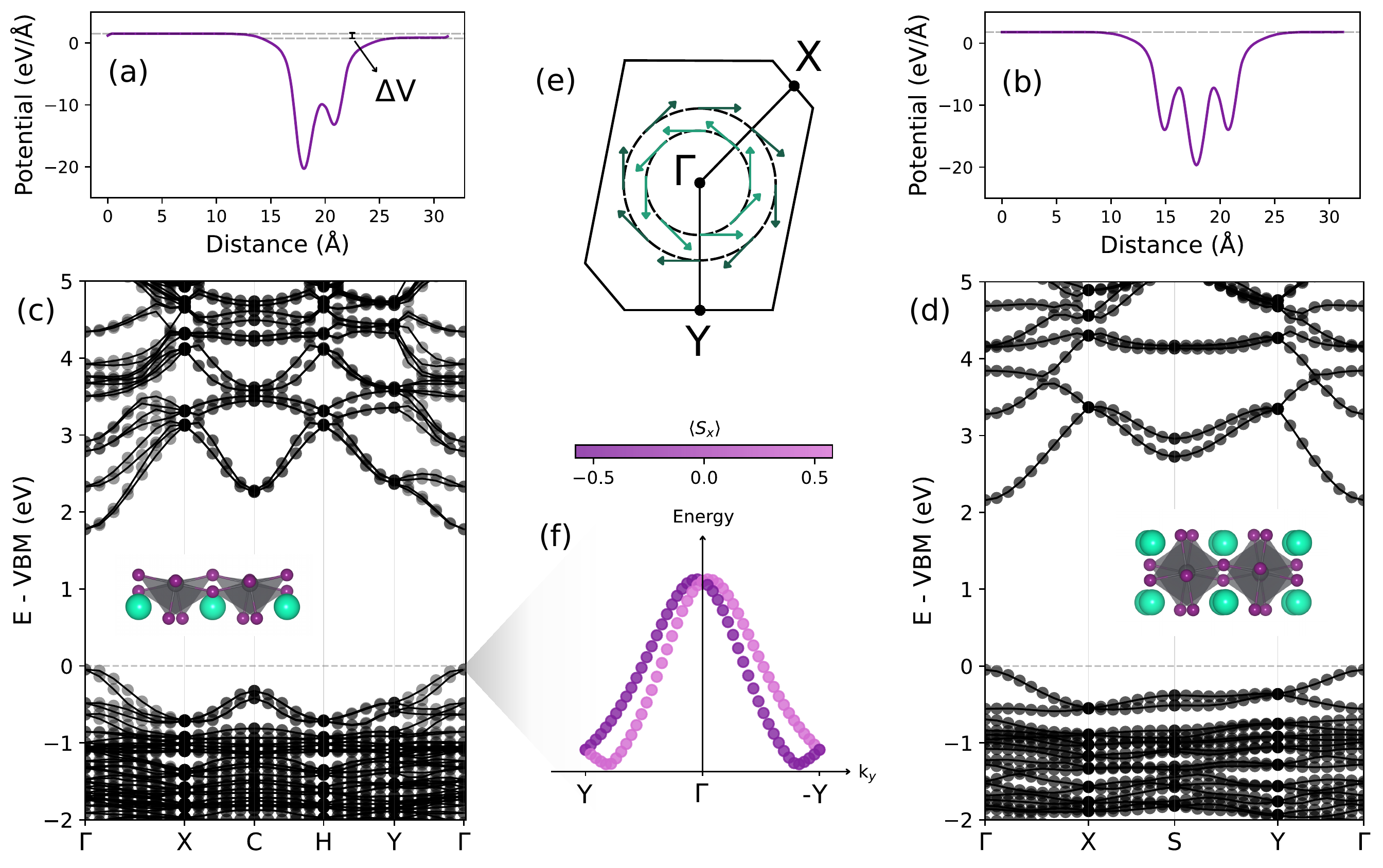}
    \caption{Planar-averaged electrostatic potential along the $z$ direction for CsPbI$_3$ and Cs$_2$PbI$_4$ stoichiometries (a--b). Electronic band structures (c--d) with HSE06 and SOC for CsPbI$_3$ and Cs$_2$PbI$_4$ stoichiometries in the P-Oblique and P-Rectangular phases, respectively. Scheme (e) shown a energy contour line relative to the CsPbI$_3$ monolayer, representing the spin orientations around the reciprocal space. Graphic (f) presents new DFT calculations, considering the SOC, for CsPbI$_3$ monolayer through the path Y $\rightarrow \Gamma \rightarrow$ -Y, confirming the Rashba-characteristic spin splitting in this monolayer.}
    \label{fig:fig5}
\end{figure*}

Regarding the comparison between the different monolayer stoichiometries, the Fig.~\ref{fig:fig5} presents some electronic properties of CsPbI$_3$ and Cs$_2$PbI$_4$, that can be generalized to the other compositions (see the SI). The planar-averaged electrostatic potentials shown in Fig.~\ref{fig:fig5}(a--b) exhibit a distinction between the ABX$_3$ and A$_2$BX$_4$ stoichiometries, namely the presence or absence of an intrinsic vertical dipole moment. Owing to their mirror symmetry with respect to the $xy$ plane, A$_2$BX$_4$ monolayers exhibit identical electrostatic potential levels at both interfaces, resulting in no internal dipole. In contrast, the ABX$_3$ monolayers lack this mirror symmetry, leading to a difference between the electrostatic potentials at opposite surfaces of the perovskene layer (denoted as $\Delta V$), which can be exploited in device architectures to control charge carrier dynamics \cite{gong_huang_yu_hu_liu_meng_wen_chen_2023}.

In terms of the electronic band dispersions in momentum space, Figs.~\ref{fig:fig5}(c--d) show the band structures for the ABX$_3$ and A$_2$BX$_4$ stoichiometries in the P-Oblique and P-Rectangular phases, respectively. Although the two systems exhibit different band gaps (with a difference of approximately $0.4$~eV), the most distinctive feature is the spin splitting observed in the ABX$_3$ stoichiometry. As illustrated in Fig.~\ref{fig:fig5}(f), the spin texture is perpendicular to the crystal momentum, a hallmark of the Rashba effect, schematically represented in Fig.~\ref{fig:fig5}(e). This spin degeneracy lifting originates from the internal dipole moment of the ABX$_3$ monolayer, oriented along the $z$ direction ($\boldsymbol{\mu} = \mu\,\boldsymbol{z}$). As a result, a Rashba effective field emerges, given by $\boldsymbol{\Omega}_R(\boldsymbol{k}) = \lambda_R\,(\boldsymbol{z} \times \boldsymbol{k})$ \cite{MeraAcosta2019Aug}. This effective field $\boldsymbol{\Omega}_R$ then couples to the spin angular momentum $\boldsymbol{S}$, leading to a splitting of the electronic states according to their alignment with the Rashba effective field.

In the Table~\ref{tab:elecprops} we summarized the electronic results, showing the band gaps and work functions for the monomorphic phases and the most stable polymorphic phases of ABX$_3$ and A$_2$BX$_4$ perovskenes. A clear trend, well known from three-dimensional halide perovskites, is observed with respect to halogen substitution: moving from I to Br to Cl leads to an increase in the band gap. This behavior can be qualitatively explained by the increasing electronegativity of the halogens, which lowers the energy of their $p$ orbitals and, consequently, the valence band maximum, resulting in a wider band gap.

On the other hand, phases exhibiting octahedral tilting systematically display larger band gaps than the M-Quadratic phase, with increases of approximately $7\%$ for ABX$_3$ and $37\%$ for A$_2$BX$_4$. This enhancement is closely related to symmetry breaking induced by the structural distortions \cite{sabino_zhao_dalpian_zunger_2024}. Table~\ref{tab:elecprops} also reports the work functions for all analyzed layers. For ABX$_3$ perovskenes, different vacuum levels associated with the distinct surface terminations lead to different work functions for the top (PbX$_2$) and bottom (CsX) surfaces. Overall, the work functions lie in the range of approximately $5$--$6$~eV and do not exhibit a clear dependence on composition or crystalline phase.

\begin{table*}
    \centering 
    \caption{Electronic properties of halide perovskite monolayers. The band gap energy ($\Delta$E$_{g}$), and the work functions ($\Phi_{top}$, and $\Phi_{bottom}$) are in eV. The M- indicates the monomorphic phase and P- indicates the polymorphic phases.} 
    \begin{tabular}{llllllllll} 
    \hline 
    Property \;\;\;\;& CsPbI$_3$ \;\; & Cs$_2$PbI$_4$ \;\; & CsPbBr$_3$ \;\; & Cs$_2$PbBr$_4$ \;\; & CsPbCl$_3$ \;\; & Cs$_2$PbCl$_4$ \\ 
    \hline 
    M-$\Delta$E$_{g}$ & 1.61 & 1.51 & 2.37 & 1.95 & 3.16 & 2.43 \\ P-$\Delta$E$_{g}$ & 1.82 & 2.21 & 2.51 & 2.70 & 3.22 & 3.13 \\ M-$\Phi_{top}$ & 5.06 & 5.01 & 5.35 & 4.97 & 5.53 & 5.35 \\ P-$\Phi_{top}$ & 5.26 & 5.05 & 5.50 & 5.15 & 5.65 & 5.12 \\ M-$\Phi_{bottom}$ & 5.75 & 5.01 & 5.95 & 4.97 & 6.04 & 5.35 \\ P-$\Phi_{bottom}$ & 5.89 & 5.05 & 6.09 & 5.15 & 6.18 & 5.12 \\ \end{tabular} 
    \label{tab:elecprops} 
\end{table*}

\begin{figure*}[h!]
    \centering
    \includegraphics[width=1\linewidth]{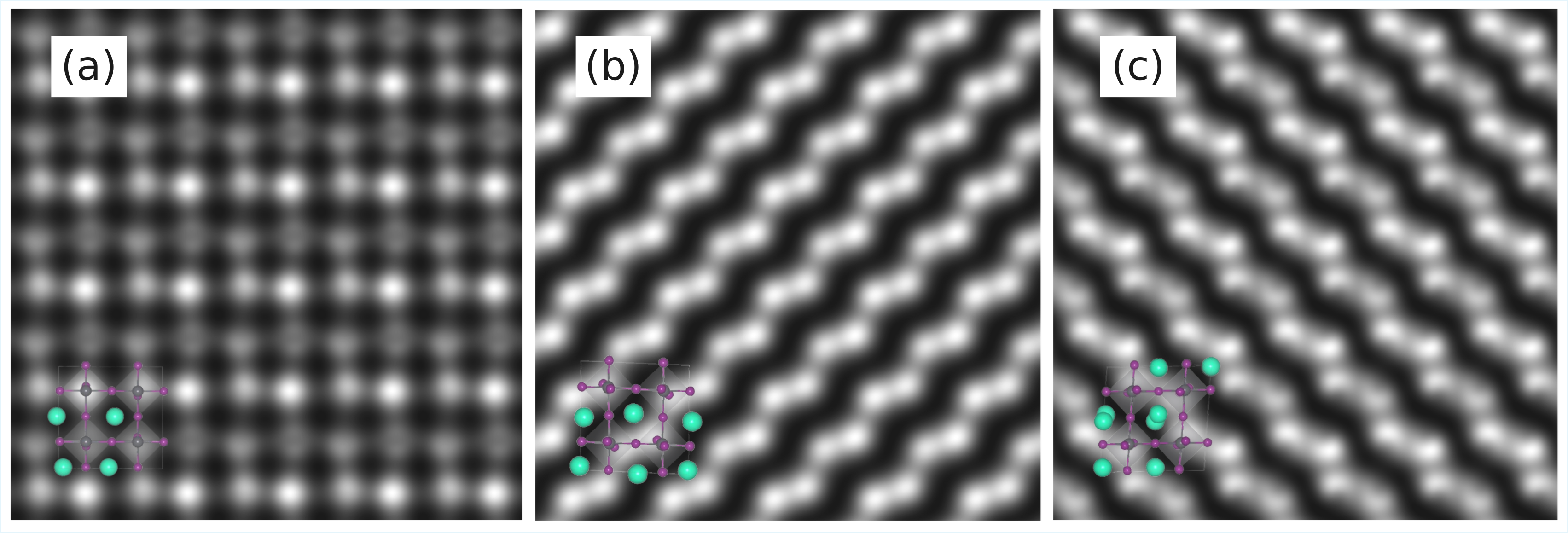}
    \caption{Simulated STM images for (a) P-Rectangular CsPbI$_3$, (b) P-Rectangular CsPbI$_4$, and (c) P-Rectangular Cs$_2$PbI$_4$ monolayers. A bias of $-0.5$~eV was applied, probing filled states.}
    \label{fig:fig7}
\end{figure*}

Finally, insight into the electronic structure also enables predictions regarding experimental identification using scanning tunneling microscopy (STM). Figure~\ref{fig:fig7} presents simulated STM images for different stoichiometries, obtained by probing the PbI$_2$, I, and CsI terminations of CsPbI$_3$, CsPbI$_4$, and Cs$_2$PbI$_4$ monolayers, respectively. In the CsPbI$_3$ monolayer [Fig.~\ref{fig:fig7}(a)], halogen atoms appear brighter due to their dominant contribution to the valence band, while Pb atoms remain visible enough to delineate the square lattice. In the CsPbI$_4$ and Cs$_2$PbI$_4$ monolayers [Figs.~\ref{fig:fig7}(b--c)], diagonal features associated with halogen dimer assemblies emerge as a consequence of surface reconstruction, consistent with experimental observations in hybrid perovskites \cite{sanjay_2020}. Although the lack of direct Cs contribution to the valence states complicates the distinction between ABX$_4$ and A$_2$BX$_4$ stoichiometries, indirect signatures related to dimer arrangement can still be identified.

From the perspective of functional materials development, the combined understanding of mechanical softness and the strong coupling between structural distortions and electronic properties in perovskenes opens promising avenues for tuning optoelectronic behavior via mechanical or epitaxial perturbations. At the same time, the low rigidity of these materials represents both a challenge for device stability and an opportunity for engineering emergent functionalities through controlled structural modifications.

\section{Conclusions}

Two-dimensional halide perovskites are promising materials, with a growing number of experimental works exploring their properties and potential applications. Their characteristics combine the remarkable optoelectronic properties of halide perovskites with the versatility of two-dimensional materials, yet a comprehensive understanding of their properties is still lacking. In this context, this work investigates, through first-principles computational simulations, the structural, elastic, and electronic behaviors of this emerging class of materials, aiming to contribute to their understanding and description.

In terms of structural properties, it was verified that the tilting of the octahedra forming the perovskene lattice is responsible for enhancing the stability of these materials. Furthermore, the existence of possible ABX$_3$ and A$_2$BX$_4$ stoichiometries was assessed under different chemical environments, while the ABX$_4$ stoichiometry was confirmed to be unstable, in the freestanding form, when compared to the other two stoichiometries. In addition, a possible way to discriminate among the different stoichiometries was proposed from the analysis of the STM simulations.

From a mechanical perspective, we described the behavior of the Young’s modulus, shear modulus, layer modulus, and Poisson’s ratio, as well as their relation to the chemical bonds within the perovskene lattice. In this analysis, we showed a decrease of approximately $50\%$ in the Young’s modulus when octahedral tilting is considered, along with an increase in Poisson’s ratio due to this dynamic.

Regarding electronic properties, some trends previously observed in 3D perovskites were also verified for perovskenes, such as the dominant contributions of Pb $s$ and $p$ orbitals to the conduction band, the major role of halogen $p$ orbitals in the valence band, and the influence of halogen electronegativity and structural symmetry breaking on the band gap. Lastly, the spin splitting presented in the ABX$_3$ monolayer emerges as an interesting feature of these materials in two-dimensionality, and shows how the internal dipole moment can generate a degeneracy breaking through the Rashba effect.

Overall, this advance in the description of the characteristics and properties of perovskenes opens a path for future experimental studies exploring the modification of these materials to tune their properties. Moreover, it establishes a foundation for new theoretical and computational works aimed at investigating other properties of this class of materials.

\section{Methods}

All first-principles calculations were performed within the framework of density functional theory (DFT) using the Vienna \textit{Ab initio} Simulation Package (VASP). A plane-wave basis set was employed with a cutoff energy of 600~eV, and interactions between core and valence electrons were described using the projector augmented-wave (PAW) method. For Cs atoms, the 5s$^2$ 5p$^6$ 6s$^1$; for Pb atoms, the 6s$^2$ 6p$^2$; and for the halogens, the ns$^2$ np$^5$ electrons were taken into account in the self-consistent steps. The exchange--correlation functional adopted was PBEsol, a generalized gradient approximation (GGA) formulation suitable for solids \cite{li_zhao_chu_gao_lv_wang_tang_hong_2022}. Electronic self-consistency was achieved with an energy convergence criterion of 10$^{-6}$~eV, while ionic relaxations were considered converged when the maximum force on each atom was below 0.01~eV/\AA. Regarding the electronic properties, hybrid-functional calculations were carried out using PBE and HSE06. Spin--orbit coupling (SOC) was included using the noncollinear PAW formalism as implemented in VASP \cite{Steiner2016Jun}.

Three different halogens were considered: Cl, Br, and I, and three primary stoichiometries were investigated: ABX$_3$, ABX$_4$, and A$_2$BX$_4$ \cite{ricciardulli_yang_smet_saliba_2021}, as shown in the Figure~\ref{fig:fig1}. For each combination of halogen and stoichiometry, different slab representations of the repetition unit were considered, leading to different in-plane symmetries and phases, as discussed in the Results section. Figure~\ref{fig:fig1} illustrates the differences between the systems.

After the initial structural relaxation using the established maximum force criterion, the different phases and stoichiometries were compared based on the formation energy ($\Delta_fE^{(2D)}$), given by $E^{(2D)} - \sum_i n_i \mu_i^{(0)}$, where $E^{(2D)}$ corresponds to the total energy of the 2D material obtained from first-principles calculations, and $n_i$ and $\mu_i^{(0)}$ represent the number of atoms of species $i$ in the system and the reference chemical potential of species $i$, respectively. The values of $\mu_i^{(0)}$ were obtained from first-principles calculations for cesium and lead in the solid phase and for halogens in the gaseous $X_2$ phase, using the same basis set, pseudopotentials, and exchange--correlation functional employed for the 2D materials.

To explore the influence of cesium- and halogen-rich or -poor environments (Cl, Br, I), the grand canonical free energy ($\Omega$) was calculated, which relates thermodynamic stability to the chemical potential of the environment \cite{reuter_scheffler_2001, li_wang_2025}. In cases where entropic contributions are small compared to the internal energy, the grand canonical free energy can be calculated using Eq.~\ref{eq:Landau}, where $\Delta \mu = \mu - \mu^{(0)}$:

\begin{equation}
    \Omega^{(2D)} = \Delta_f E^{(2D)} - \sum_i n_i \Delta \mu_i
    \label{eq:Landau}
\end{equation}

As the upper limit for the chemical potential of cesium and halogens ($\mu_{\mathrm{Cs}}, \mu_X$), the standard chemical potentials ($\mu_{\mathrm{Cs}}^{(0)}, \mu_X^{(0)}$) were considered, while the lower limit was defined by the condition $\Omega^{(2D)} (\Delta \mu_{\mathrm{Cs}}, \Delta \mu_X) = 0$, corresponding to the onset of thermodynamic instability of the material. The grand canonical free energy was then used as a criterion to construct the phase diagrams of the three stoichiometries as a function of ($\Delta \mu_{\mathrm{Cs}}, \Delta \mu_X$).

In addition to thermodynamic stability, the mechanical properties were also assessed from first principles through the calculation of the elastic tensor of the 2D materials. For this purpose, two methods were evaluated: the finite-difference and energy--strain approaches \cite{li_zhu_wang_pan_cao_wu_chen_2023}. The finite-difference method, implemented directly in VASP, determines the elastic tensor by performing six finite lattice distortions and deriving the elastic constants from the stress--strain relationship \cite{yvon_2002, wu_2005}. In this case, to convert the three-dimensional elastic tensor into the two-dimensional tensor format, the Python library \textit{mechelastic} \cite{singh_2021} was also employed.

For the energy--strain method, the \textit{VASPKIT} implementation was used, which generates deformations corresponding to the tensor elements according to the two-dimensional symmetry of the materials and calculates each element from the relationship between the total energies and the respective deformations \cite{wang_xu_liu_tang_geng_2021}. Additionally, the equations used to calculate the mechanical properties from the elastic tensor are detailed in the Supplementary Information (SI), as well as the mechanical stability conditions for 2D materials \cite{born1988dynamical}.


\bibliographystyle{unsrt} 
\bibliography{references}

\begin{thebibliography}{10}

\bibitem{dong_ran_gao_li_xia_huang_2023}
He~Dong, Chenxin Ran, Weiyin Gao, Mingjie Li, Yingdong Xia, and Wei Huang.
\newblock Metal halide perovskite for next-generation optoelectronics: Progresses and prospects.
\newblock {\em eLight}, 3(1), 2023.

\bibitem{manser_christians_kamat_2016}
Joseph~S. Manser, Jeffrey~A. Christians, and Prashant~V. Kamat.
\newblock Intriguing optoelectronic properties of metal halide perovskites.
\newblock {\em Chem. Rev.}, 116(21):12956--13008, 2016.

\bibitem{jena_2019}
Ajay~Kumar Jena, Ashish Kulkarni, and Tsutomu Miyasaka.
\newblock Halide perovskite photovoltaics: Background, status, and future prospects.
\newblock {\em Chem. Rev.}, 119(5):3036--3103, 2019.

\bibitem{liu_ma_zhang_he_dai_li_shi_manna_2025}
Ying Liu, Zhuangzhuang Ma, Jibin Zhang, Yanni He, Jinfei Dai, Xinjian Li, Zhifeng Shi, and Liberato Manna.
\newblock Light-emitting diodes based on metal halide perovskite and perovskite-related nanocrystals.
\newblock {\em Adv. Mater.}, 2025.

\bibitem{huang_yu_wu_li_li_2024}
Yajie Huang, Jiaxing Yu, Zhiyuan Wu, Borui Li, and Ming Li.
\newblock All-inorganic lead halide perovskites for photocatalysis: A review.
\newblock {\em RSC Adv.}, 14(7):4946--4965, 2024.

\bibitem{morais_caturello_lemes_ferreira_ferreira_acuña_brochsztain_dalpian_souza_2024}
Eliane~A. Morais, Naidel A. M.~S. Caturello, Maykon~A. Lemes, Henrique Ferreira, Fabio~F. Ferreira, Jose J.~S. Acu{\~n}a, Sergio Brochsztain, Gustavo~M. Dalpian, and Jose~A. Souza.
\newblock Rashba spin splitting limiting the application of 2d halide perovskites for uv-emitting devices.
\newblock {\em ACS Appl. Mater. Interfaces}, 16(3):4261--4270, 2024.

\bibitem{zheng_pauporté_2023}
Daming Zheng and Thierry Pauport{\'e}.
\newblock Advances in optical imaging and optical communications based on high-quality halide perovskite photodetectors.
\newblock {\em Adv. Funct. Mater.}, 34(11), 2023.

\bibitem{shamim_2020}
Shamim Shahrokhi, Wenxiu Gao, Yutao Wang, Pradeep~Raja Anandan, Md.~Zahidur Rahaman, Simrjit Singh, Danyang Wang, Claudio Cazorla, Guoliang Yuan, Jun-Ming Liu, and Tom Wu.
\newblock Emergence of ferroelectricity in halide perovskites.
\newblock {\em Small Methods}, 4(8), 2020.

\bibitem{zheng_2022}
Weilin Zheng, Xiucai Wang, Xin Zhang, Bing Chen, Hao Suo, Zhifeng Xing, Yanze Wang, Han-Lin Wei, Jiangkun Chen, Yang Guo, and Feng Wang.
\newblock Emerging halide perovskite ferroelectrics.
\newblock {\em Adv. Mater.}, 35(21), 2022.

\bibitem{jin_im_freeman_2012}
Hosub Jin, Jino Im, and Arthur~J. Freeman.
\newblock Topological insulator phase in halide perovskite structures.
\newblock {\em Phys. Rev. B}, 86(12), 2012.

\bibitem{lafuente-bartolome_lian_giustino_2024}
Jon Lafuente-Bartolome, Chao Lian, and Feliciano Giustino.
\newblock Topological polarons in halide perovskites.
\newblock {\em Proc. Natl. Acad. Sci. U.S.A.}, 121(21), 2024.

\bibitem{kim_han_choi_kim_jang_2018}
Hyojung Kim, Ji~Su Han, Jaeho Choi, Soo~Young Kim, and Ho~Won Jang.
\newblock Halide perovskites for applications beyond photovoltaics.
\newblock {\em Small Methods}, 2(3), 2018.

\bibitem{akkerman_manna_2020}
Quinten~A. Akkerman and Liberato Manna.
\newblock What defines a halide perovskite?
\newblock {\em ACS Energy Lett.}, 5(2):604--610, 2020.

\bibitem{wang_wang_doherty_stranks_gao_yang_2025}
Yong Wang, Yu~Wang, Tiarnan A.~S. Doherty, Samuel~D. Stranks, Feng Gao, and Deren Yang.
\newblock Octahedral units in halide perovskites.
\newblock {\em Nat. Rev. Chem.}, 9(4):261--277, 2025.

\bibitem{zhao_dalpian_wang_zunger_2020}
Xin-Gang Zhao, Gustavo~M. Dalpian, Zhi Wang, and Alex Zunger.
\newblock Polymorphous nature of cubic halide perovskites.
\newblock {\em Phys. Rev. B}, 101(15), 2020.

\bibitem{hautzinger_mihalyi-koch_jin_2024}
Matthew~P. Hautzinger, Willa Mihalyi-Koch, and Song Jin.
\newblock A-site cation chemistry in halide perovskites.
\newblock {\em Chem. Mater.}, 36(21):10408--10420, 2024.

\bibitem{chu_chu_zhao_ye_jiang_zhang_you_2021}
Zema Chu, Xinbo Chu, Yang Zhao, Qiufeng Ye, Ji~Jiang, Xingwang Zhang, and Jingbi You.
\newblock Emerging low-dimensional crystal structure of metal halide perovskite optoelectronic materials and devices.
\newblock {\em Small Struct.}, 2(6), 2021.

\bibitem{Leite2026Oct}
Marcos de~C. Leite, Gabriel~X. Pereira, Lucas~M. Farigliano, Gustavo~M. Dalpian, Juan Andr{\'e}s, and Amanda~F. Gouveia.
\newblock Mapping and characterization of surface-dependent electronic properties and morphological changes in cubic crystalline perovskite {CsPbBr3}.
\newblock {\em Comput. Mater. Sci.}, 264:114477, 2026.

\bibitem{kim_park_cho_jang_2024}
Seung~Ju Kim, Sungwoo Park, Hyo~Min Cho, and Ho~Won Jang.
\newblock Low-dimensional halide perovskites for advanced electronics.
\newblock {\em Mater. Today Electron.}, 9:100111, 2024.

\bibitem{rafique_abbas_mendes_barquinha_martins_fortunato_águas_jana_2025}
Humaira Rafique, Ghulam Abbas, Manuel~J. Mendes, Pedro Barquinha, Rodrigo Martins, Elvira Fortunato, Hugo {\'A}guas, and Santanu Jana.
\newblock Recent advancements and perspectives of low-dimensional halide perovskites for visual perception and optoelectronic applications.
\newblock {\em Nano-Micro Lett.}, 18(1), 2025.

\bibitem{laxmi_2025}
Laxmi, Phalguna Srinivasan, and Dinesh Kabra.
\newblock Optical and optoelectronic properties of 2d, quasi-2d, and 3d metal halide perovskites.
\newblock {\em J. Mater. Chem. C}, 2025.

\bibitem{D3CP04435A}
Mosayeb Naseri, Shirin Amirian, Mehrdad Faraji, Mohammad~Abdur Rashid, Maicon~Pierre Louren{\c{c}}o, Venkataraman Thangadurai, and D.~R. Salahub.
\newblock Perovskenes: Two-dimensional perovskite-type monolayer materials predicted by first-principles calculations.
\newblock {\em Phys. Chem. Chem. Phys.}, 26(2):946--957, 2024.

\bibitem{FARHADI2025113581}
Bita Farhadi, Hatef Shahmohamadi, Zhen Xu, Jingjing Zhang, Tang Tang, Mosayeb Naseri, Sergey Gusarov, and Dennis~R. Salahub.
\newblock Monolayer {Ba2TiO4} and {Sr2TiO4} perovskenes: A first-principles study.
\newblock {\em Mater. Today Commun.}, 48:113581, 2025.

\bibitem{mao_stoumpos_kanatzidis_2018}
Lingling Mao, Constantinos~C. Stoumpos, and Mercouri~G. Kanatzidis.
\newblock Two-dimensional hybrid halide perovskites: Principles and promises.
\newblock {\em J. Am. Chem. Soc.}, 141(3):1171--1190, 2018.

\bibitem{wu_liang_zhang_ge_xing_sun_2021}
Guangbao Wu, Rui Liang, Zhipeng Zhang, Mingzheng Ge, Guichuan Xing, and Guoxing Sun.
\newblock 2d hybrid halide perovskites: Structure, properties, and applications in solar cells.
\newblock {\em Small}, 17(43), 2021.

\bibitem{milot_2016}
Rebecca~L. Milot, Rebecca~J. Sutton, Giles~E. Eperon, Amir~Abbas Haghighirad, Josue Martinez~Hardigree, Laura Miranda, Henry~J. Snaith, Michael~B. Johnston, and Laura~M. Herz.
\newblock Charge-carrier dynamics in 2d hybrid metal--halide perovskites.
\newblock {\em Nano Lett.}, 16(11):7001--7007, 2016.

\bibitem{ricciardulli_yang_smet_saliba_2021}
Antonio~Gaetano Ricciardulli, Sheng Yang, Jurgen~H. Smet, and Michael Saliba.
\newblock Emerging perovskite monolayers.
\newblock {\em Nat. Mater.}, 20(10):1325--1336, 2021.

\bibitem{thakur_chang_2023}
Diksha Thakur and Sheng~Hsiung Chang.
\newblock Material properties and optoelectronic applications of lead halide perovskite thin films.
\newblock {\em Synth. Met.}, 301:117535, 2023.

\bibitem{orange_tambwe_arendse_malevu_ross_2024}
Tamsen Orange, Kevin Tambwe, Christopher Arendse, Thembinkosi Malevu, and Natasha Ross.
\newblock Constructing hybrid semiconductor thin-films for advanced photovoltaics.
\newblock {\em ChemistrySelect}, 9(8), 2024.

\bibitem{li_zhu_wang_pan_cao_wu_chen_2023}
Mingguang Li, Zheng Zhu, Zhizhi Wang, Wenjing Pan, Xinxiu Cao, Guangbao Wu, and Runfeng Chen.
\newblock High-quality hybrid perovskite thin films by post-treatment technologies in photovoltaic applications.
\newblock {\em Adv. Mater.}, 36(7), 2023.

\bibitem{yehonadav_2015}
Yehonadav Bekenstein, Brent~A. Koscher, Samuel~W. Eaton, Peidong Yang, and A.~Paul Alivisatos.
\newblock Highly luminescent colloidal nanoplates of perovskite cesium lead halide and their oriented assemblies.
\newblock {\em J. Am. Chem. Soc.}, 137(51):16008--16011, 2015.

\bibitem{akkerman_motti_2016}
Quinten~A. Akkerman, Silvia~Genaro Motti, Ajay Ram, Edoardo Mosconi, Valerio D'Innocenzo, Giovanni Bertoni, Sergio Marras, Brett~A. Kamino, Laura Miranda, Filippo De~Angelis, Annamaria Petrozza, Mirko Prato, and Liberato Manna.
\newblock Solution synthesis approach to colloidal cesium lead halide perovskite nanoplatelets with monolayer-level thickness control.
\newblock {\em J. Am. Chem. Soc.}, 138(3):1010--1016, 2016.

\bibitem{niu_2014}
Wendy Niu, Anna Eiden, G.~Vijaya Prakash, and Jeremy~J. Baumberg.
\newblock Exfoliation of self-assembled 2d organic--inorganic perovskite semiconductors.
\newblock {\em Appl. Phys. Lett.}, 104(17), 2014.

\bibitem{yaffe_2015}
Omer Yaffe, Alexey Chernikov, Zachariah~M. Norman, Yu~Zhong, A.~Velauthapillai, Jonathan~S. Owen, and Tony~F. Heinz.
\newblock Excitons in ultrathin organic--inorganic perovskite crystals.
\newblock {\em Phys. Rev. B}, 92(4), 2015.

\bibitem{weidman_seitz_stranks_tisdale_2016}
Mark~C. Weidman, Michael Seitz, Samuel~D. Stranks, and William~A. Tisdale.
\newblock Highly tunable colloidal perovskite nanoplatelets through variable cation, metal, and halide composition.
\newblock {\em ACS Nano}, 10(8):7830--7839, 2016.

\bibitem{hu_2024}
Shenghan Hu, Peiran Hou, Changyu Duan, Yichen Dou, Xinyu Deng, Wenjuan Xiong, Zhangwei Yuan, Jiace Liang, Yong Peng, Yi-Bing Cheng, and Zhiliang Ku.
\newblock Vapor--solid reaction techniques for the growth of organic--inorganic hybrid perovskite thin films.
\newblock {\em Small}, 21(6), 2024.

\bibitem{dou_2015}
Letian Dou, Andrew~B. Wong, Yi~Yu, Minliang Lai, Nikolay Kornienko, Samuel~W. Eaton, Anthony Fu, Connor~G. Bischak, Jie Ma, Tina Ding, Naomi~S. Ginsberg, Lin-Wang Wang, A.~Paul Alivisatos, and Peidong Yang.
\newblock Atomically thin two-dimensional organic--inorganic hybrid perovskites.
\newblock {\em Science}, 349(6255):1518--1521, 2015.

\bibitem{leng_2018}
Kai Leng, Ibrahim Abdelwahab, Ivan Verzhbitskiy, Mykola Telychko, Leiqiang Chu, Wei Fu, Xiao Chi, Na~Guo, Zhihui Chen, Zhongxin Chen, Chun Zhang, Qing-Hua Xu, Jiong Lu, Manish Chhowalla, Goki Eda, and Kian~Ping Loh.
\newblock Molecularly thin two-dimensional hybrid perovskites with tunable optoelectronic properties due to reversible surface relaxation.
\newblock {\em Nat. Mater.}, 17(10):908--914, 2018.

\bibitem{shi_2020}
Enzheng Shi, Biao Yuan, Stephen~B. Shiring, Yao Gao, Akriti, Yunfan Guo, Cong Su, Minliang Lai, Peidong Yang, Jing Kong, Brett~M. Savoie, Yi~Yu, and Letian Dou.
\newblock Two-dimensional halide perovskite lateral epitaxial heterostructures.
\newblock {\em Nature}, 580(7805):614--620, 2020.

\bibitem{yang_li_2022}
Jack Yang and Sean Li.
\newblock Computational material database of free-standing 2d perovskites.
\newblock {\em ChemRxiv}, 2022.

\bibitem{ming_yang_zeng_zhang_sun_2020}
Chen Ming, Ke~Yang, Hao Zeng, Shengbai Zhang, and Yi-Yang Sun.
\newblock Octahedron rotation evolution in 2d perovskites and its impact on optoelectronic properties: The case of ba--zr--s chalcogenides.
\newblock {\em Mater. Horiz.}, 7(11):2985--2993, 2020.

\bibitem{zhang_xie_he_ji_shen_2024}
Junting Zhang, Yu~Xie, Jun He, Ke~Ji, and Xiaofan Shen.
\newblock Ferroelectricity induced by lone-pair electron effect in halide perovskite monolayers.
\newblock {\em Phys. Rev. B}, 110(3), 2024.

\bibitem{zhang_ji_he_xie_2024}
Junting Zhang, Ke~Ji, Jun He, and Yu~Xie.
\newblock Two-dimensional ferroelectricity in all-inorganic halide perovskite monolayer {Cs2PbX4} ({X} = {F}, {Cl}, {Br}, {I}).
\newblock {\em Phys. Rev. B}, 110(19), 2024.

\bibitem{lee_bristowe_lee_lee_bristowe_cheetham_jang_2016}
Jung-Hoon Lee, Nicholas~C. Bristowe, June~Ho Lee, Sung-Hoon Lee, Paul~D. Bristowe, Anthony~K. Cheetham, and Hyun~Myung Jang.
\newblock Resolving the physical origin of octahedral tilting in halide perovskites.
\newblock {\em Chemistry of Materials}, 28(12):4259--4266, May 2016.

\bibitem{farigliano_ribeiro_dalpian_2024}
Lucas~Martin Farigliano, Fabio~Negreiros Ribeiro, and Gustavo~Martini Dalpian.
\newblock Phase transitions in cspbbr3: Evaluating perovskite behavior over different time scales.
\newblock {\em Materials Advances}, 5(14):5794--5801, January 2024.

\bibitem{bonadio_2021}
A.~Bonadio, C.~A. Escanhoela, F.~P. Sabino, G.~Sombrio, V.~G. de~Paula, F.~F. Ferreira, A.~Janotti, G.~M. Dalpian, and J.~A. Souza.
\newblock Entropy-driven stabilization of the cubic phase of mapbi3 at room temperature.
\newblock {\em Journal of Materials Chemistry A}, 9(2):1089--1099, 2021.

\bibitem{Ferreira2018Aug}
A.~C. Ferreira, A.~L{\'e}toublon, S.~Paofai, S.~Raymond, C.~Ecolivet, B.~Ruffl{\'e}, S.~Cordier, C.~Katan, M.~I. Saidaminov, A.~A. Zhumekenov, O.~M. Bakr, J.~Even, and P.~Bourges.
\newblock Elastic softness of hybrid lead halide perovskites.
\newblock {\em Phys. Rev. Lett.}, 121(8):085502, 2018.

\bibitem{born1988dynamical}
M.~Born and K.~Huang.
\newblock {\em Dynamical Theory of Crystal Lattices}.
\newblock Clarendon Press, 1988.

\bibitem{sun_isikgor_deng_wei_kieslich_bristowe_ouyang_cheetham_2017}
Shijing Sun, Furkan~H. Isikgor, Zeyu Deng, Fengxia Wei, Gregor Kieslich, Paul~D. Bristowe, Jianyong Ouyang, and Anthony~K. Cheetham.
\newblock Factors influencing the mechanical properties of formamidinium lead halides and related hybrid perovskites.
\newblock {\em ChemSusChem}, 10(19):3740--3745, June 2017.

\bibitem{walsh_2015}
Aron Walsh.
\newblock Principles of chemical bonding and band gap engineering in hybrid organic--inorganic halide perovskites.
\newblock {\em The Journal of Physical Chemistry C}, 119(11):5755--5760, February 2015.

\bibitem{min_2023}
Seonhong Min, Hyejin Choe, Hyun~Jung Seo, and Junsang Cho.
\newblock How chemical bonding impacts halide perovskite nanocrystals growth to bulk films: Implication of pb--x bond on growth kinetics.
\newblock {\em ChemPhysChem}, 24(14), May 2023.

\bibitem{jiang_zheng_liu_wang_2019}
Hanjun Jiang, Lu~Zheng, Zheng Liu, and Xuewen Wang.
\newblock Two-dimensional materials: From mechanical properties to flexible mechanical sensors.
\newblock {\em InfoMat}, 2(6):1077--1094, December 2019.

\bibitem{chen_liu_li_cheng_ma_wang_li_2020}
Yingying Chen, Zeyi Liu, Junze Li, Xue Cheng, Jiaqi Ma, Haizhen Wang, and Dehui Li.
\newblock Robust interlayer coupling in two-dimensional perovskite/monolayer transition metal dichalcogenide heterostructures.
\newblock {\em ACS Nano}, 14(8):10258--10264, July 2020.

\bibitem{zhang_2024}
Shuchen Zhang, Linrui Jin, Yuan Lu, Linghai Zhang, Jiaqi Yang, Qiuchen Zhao, Dewei Sun, Joshua J.~P. Thompson, Biao Yuan, Ke~Ma, Akriti, Jee~Yung Park, Yoon~Ho Lee, Zitang Wei, Blake~P. Finkenauer, Daria~D. Blach, Sarath Kumar, Hailin Peng, Arun Mannodi-Kanakkithodi, and Yi~Yu.
\newblock Moir{\'e} superlattices in twisted two-dimensional halide perovskites.
\newblock {\em Nature Materials}, 23(9):1222--1229, June 2024.

\bibitem{Araujo2025Dec}
Augusto~L. Ara{\'u}jo, Pedro~H. Sophia, F.~Crasto~de Lima, and Adalberto Fazzio.
\newblock A high-throughput framework and database for twisted 2d van der waals bilayers.
\newblock {\em npj Comput. Mater.}, 12(40):40, 2025.

\bibitem{tang_2021}
Gang Tang, Philippe Ghosez, and Jiawang Hong.
\newblock Band-edge orbital engineering of perovskite semiconductors for optoelectronic applications.
\newblock {\em The Journal of Physical Chemistry Letters}, 12(17):4227--4239, April 2021.

\bibitem{gong_huang_yu_hu_liu_meng_wen_chen_2023}
Shaokuan Gong, Yuling Huang, Xuemeng Yu, Qiushi Hu, Jingjing Liu, Jiazhi Meng, Yifan Wen, and Xihan Chen.
\newblock Ultrafast dynamics in perovskite-based optoelectronic devices.
\newblock {\em Cell Reports Physical Science}, 4(9):101580, September 2023.

\bibitem{MeraAcosta2019Aug}
Carlos Mera~Acosta, Adalberto Fazzio, and Gustavo~M. Dalpian.
\newblock Zeeman-type spin splitting in nonmagnetic three-dimensional compounds.
\newblock {\em npj Quantum Materials}, 4(41):41, August 2019.

\bibitem{sabino_zhao_dalpian_zunger_2024}
Fernando~P. Sabino, Xin-Gang Zhao, Gustavo~M. Dalpian, and Alex Zunger.
\newblock Impact of symmetry breaking and spin-orbit coupling on the band gap of halide perovskites.
\newblock {\em Physical Review B}, 110(3), July 2024.

\bibitem{sanjay_2020}
Sanjay Sahare, Prachi Ghoderao, Sadaf~Bashir Khan, Yue Chan, and Shern-Long Lee.
\newblock Recent progress in hybrid perovskite solar cells through scanning tunneling microscopy and spectroscopy.
\newblock {\em Nanoscale}, 12(30):15970--15992, January 2020.

\bibitem{li_zhao_chu_gao_lv_wang_tang_hong_2022}
Shuang Li, Shenggui Zhao, Huiqi Chu, Yue Gao, Peng Lv, Vei Wang, Gang Tang, and Jiawang Hong.
\newblock Unraveling the factors affecting the mechanical properties of halide perovskites from first-principles calculations.
\newblock {\em J. Phys. Chem. C}, 126(9):4715--4725, 2022.

\bibitem{Steiner2016Jun}
Soner Steiner, Sergii Khmelevskyi, Martijn Marsmann, and Georg Kresse.
\newblock Calculation of the magnetic anisotropy with projected-augmented-wave methodology and the case study of disordered {Fe$_{1-x}$Co$_x$} alloys.
\newblock {\em Phys. Rev. B}, 93(22):224425, 2016.

\bibitem{reuter_scheffler_2001}
Karsten Reuter and Matthias Scheffler.
\newblock Composition, structure, and stability of {RuO2}(110) as a function of oxygen pressure.
\newblock {\em Phys. Rev. B}, 65(3), 2001.

\bibitem{li_wang_2025}
Kejia Li and Mengen Wang.
\newblock Density functional theory study of surface stability and phase diagram of orthorhombic {CsPbI3}.
\newblock {\em J. Phys. Chem. C}, 2025.

\bibitem{yvon_2002}
Yvon Le~Page and Paul Saxe.
\newblock Symmetry-general least-squares extraction of elastic data for strained materials from *ab initio* calculations of stress.
\newblock {\em Phys. Rev. B}, 65(10), 2002.

\bibitem{wu_2005}
Xifan Wu, David Vanderbilt, and D.~R. Hamann.
\newblock Systematic treatment of displacements, strains, and electric fields in density-functional perturbation theory.
\newblock {\em Phys. Rev. B}, 72(3), 2005.

\bibitem{singh_2021}
Sobhit Singh, Logan Lang, Viviana Dovale-Farelo, Uthpala Herath, Pedram Tavadze, Fran{\c{c}}ois-Xavier Coudert, and Aldo~H. Romero.
\newblock {MechElastic}: A python library for analysis of mechanical and elastic properties of bulk and 2d materials.
\newblock {\em Comput. Phys. Commun.}, 267:108068, 2021.

\bibitem{wang_xu_liu_tang_geng_2021}
Vei Wang, Nan Xu, Jin-Cheng Liu, Gang Tang, and Wen-Tong Geng.
\newblock {VASPKIT}: A user-friendly interface facilitating high-throughput computing and analysis using {VASP}.
\newblock {\em Comput. Phys. Commun.}, 267:108033, 2021.

\end{thebibliography}

\end{document}


\newcommand*{\getTitleGer}{Supporting Information: Intrinsic Instabilities and Mechanical Anisotropy in Halide Perovskite Monolayers}
\newcommand*{\getFaculty}{\textit{$^{a}$~Departamento de Física dos Materiais e Mecânica, Instituto de Física, Universidade de São Paulo, São Paulo 05508-090, São Paulo, Brazil.} \newline
\textit{$^{b}$~INFIQC, CONICET, Departamento de Química Teórica y Computacional, Facultad de Ciencias Químicas, Universidad Nacional de Córdoba, Argentina.}
\newline
\textit{$^{c}$~Departamento de Física dos Materiais e Mecânica, Instituto de Física, Universidade de São Paulo, São Paulo 05508-090, São Paulo, Brazil.}}
\newcommand*{\getAuthor}{Gabriel X. Pereira,\textit{$^{a}$} Lucas M. Farigliano,\textit{$^{a}$}\textit{$^{b}$} Roberto H. Miwa,\textit{$^{c}$}, and Gustavo M. Dalpian\textit{$^{\ast a}$}}
\newcommand*{\getContact}{Contact}

\begin{titlepage}
    \oddsidemargin=\evensidemargin\relax
    \textwidth=\dimexpr\paperwidth-2\evensidemargin-2in\relax
    \hsize=\textwidth\relax
    
    \centering

    \line(1,0){400}

    \vspace{0,5cm}
    
    {\huge \getTitleGer{}}

    \vspace{1cm}
    
    \getAuthor{}
    \vspace{0,25cm}

    \getFaculty{}
    \vspace{0,25cm}
  
\end{titlepage}

\tableofcontents
\clearpage

\section{Thermodynamical properties}

\begin{figure}[h]
    \centering
    \includegraphics[width=1\linewidth]{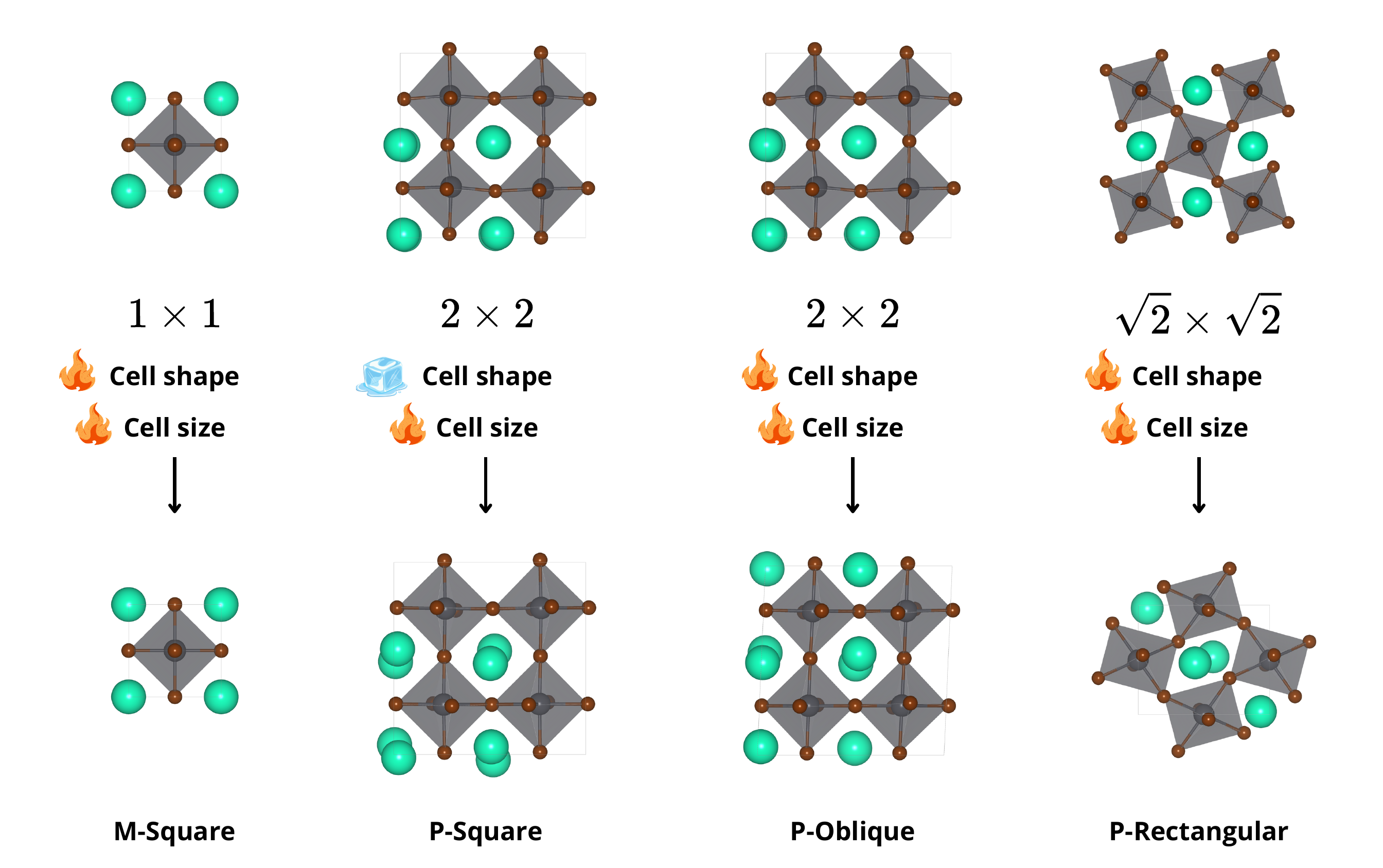}
    \caption{Summary of how each phase was generated. The fire icon means that the parameter were optmizated during the ionic relaxation, while the ice icon means that the parameter were not.}
    \label{fig:thermo1}
\end{figure}

\begin{figure}[h]
    \centering
    \includegraphics[width=.6\linewidth]{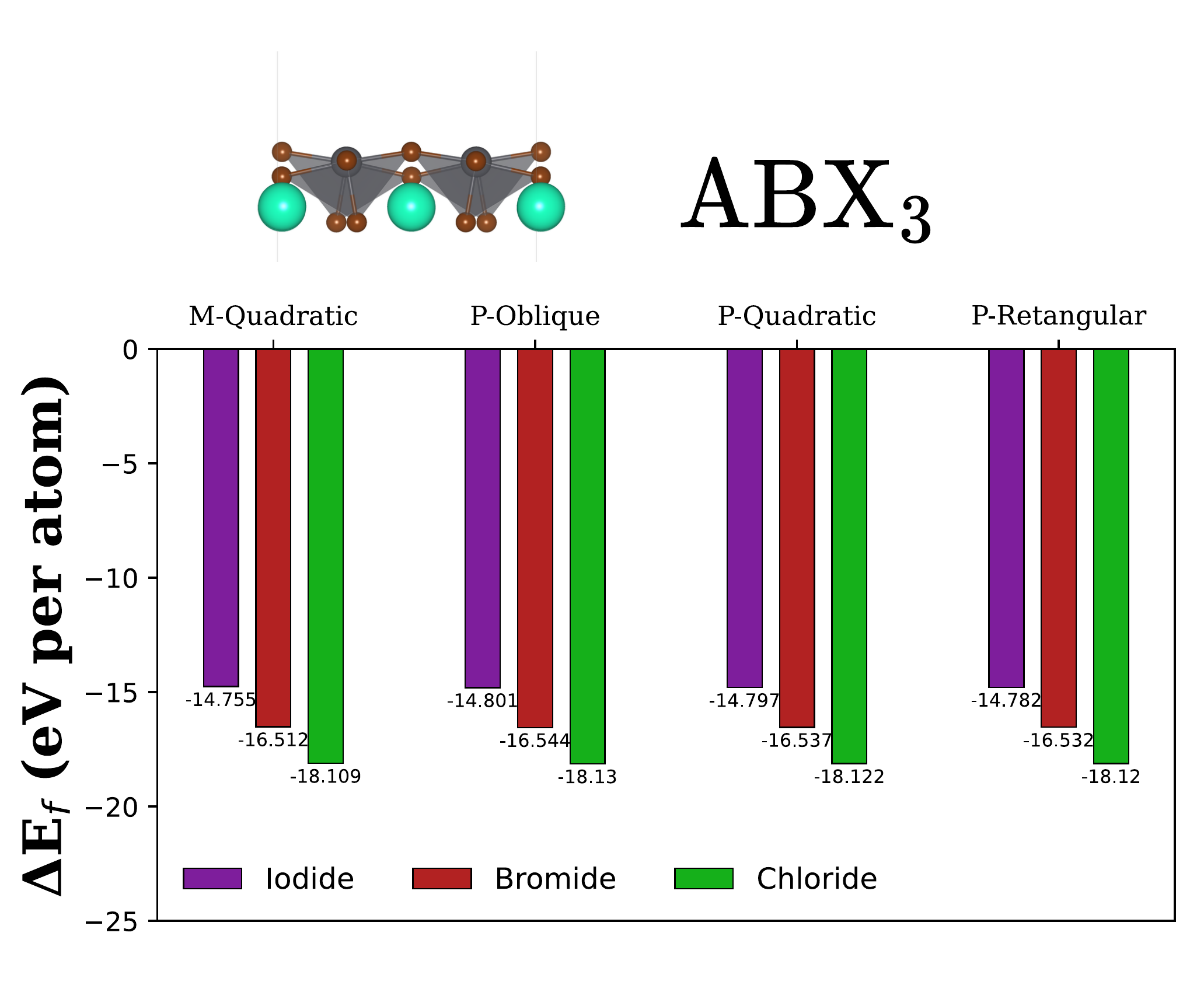}
    \caption{Formation energies for CsPbX$_3$ monolayers ($X=$ I, Br, Cl), comparing the diferent phases.}
    \label{fig:thermo2}
\end{figure}

\begin{figure}[h]
    \centering
    \includegraphics[width=.6\linewidth]{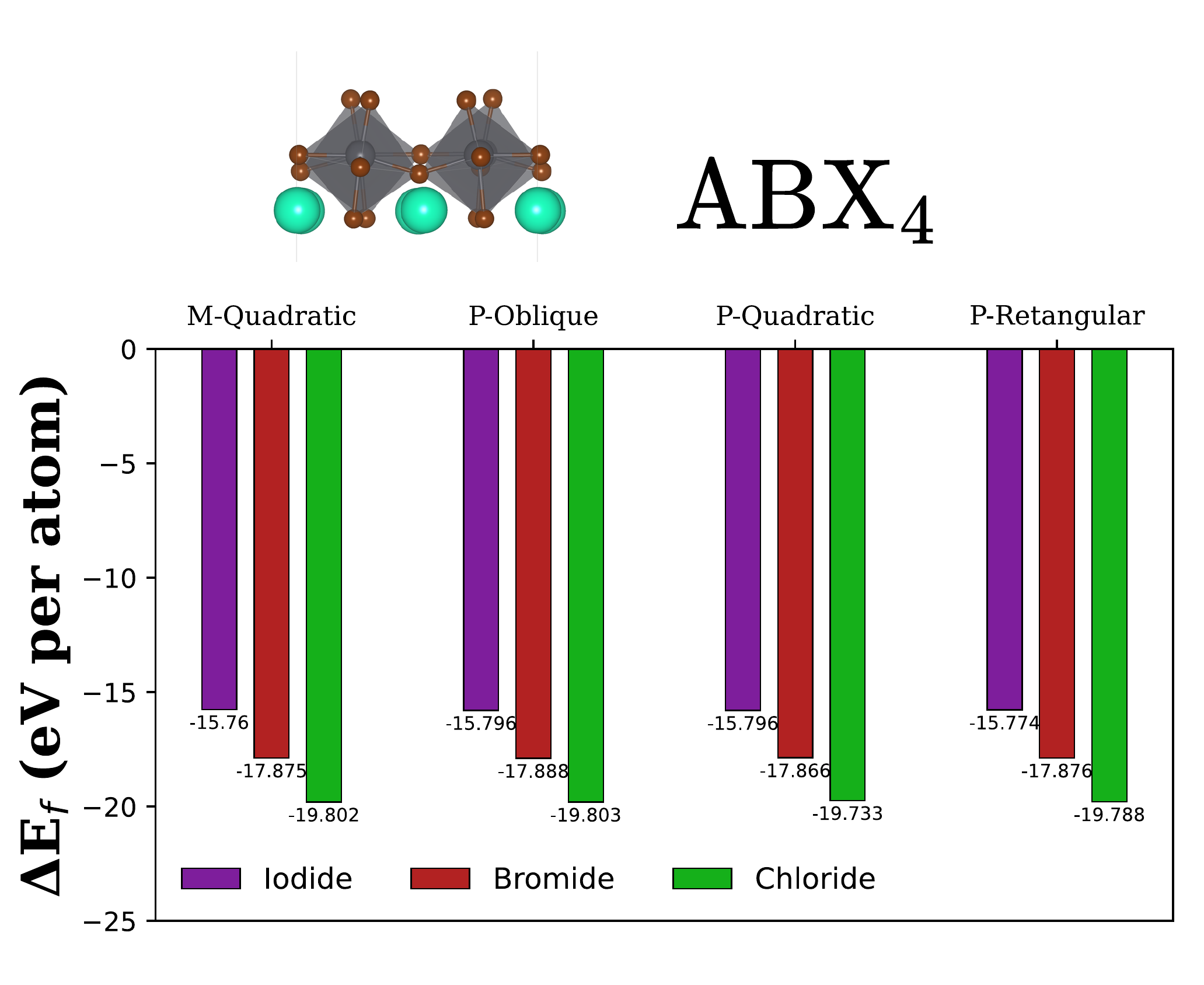}
    \caption{Formation energies for CsPbX$_4$ monolayers ($X=$ I, Br, Cl), comparing the diferent phases.}
    \label{fig:thermo3}
\end{figure}

\begin{figure}[h]
    \centering
    \includegraphics[width=.6\linewidth]{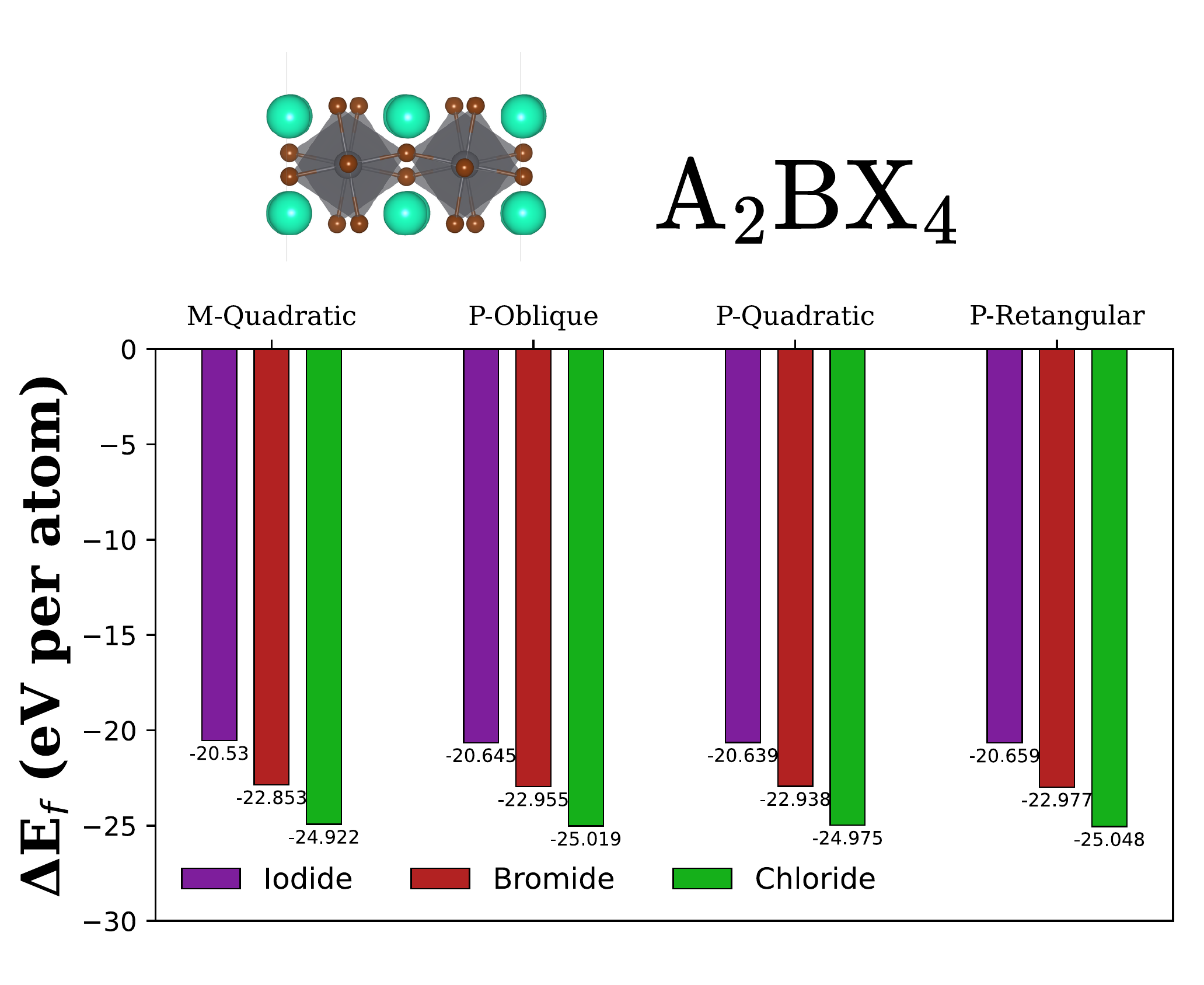}
    \caption{Formation energies for Cs$_2$PbX$_4$ monolayers ($X=$ I, Br, Cl), comparing the diferent phases.}
    \label{fig:thermo4}
\end{figure}

\begin{figure}[h]
    \centering
    \includegraphics[width=1\linewidth]{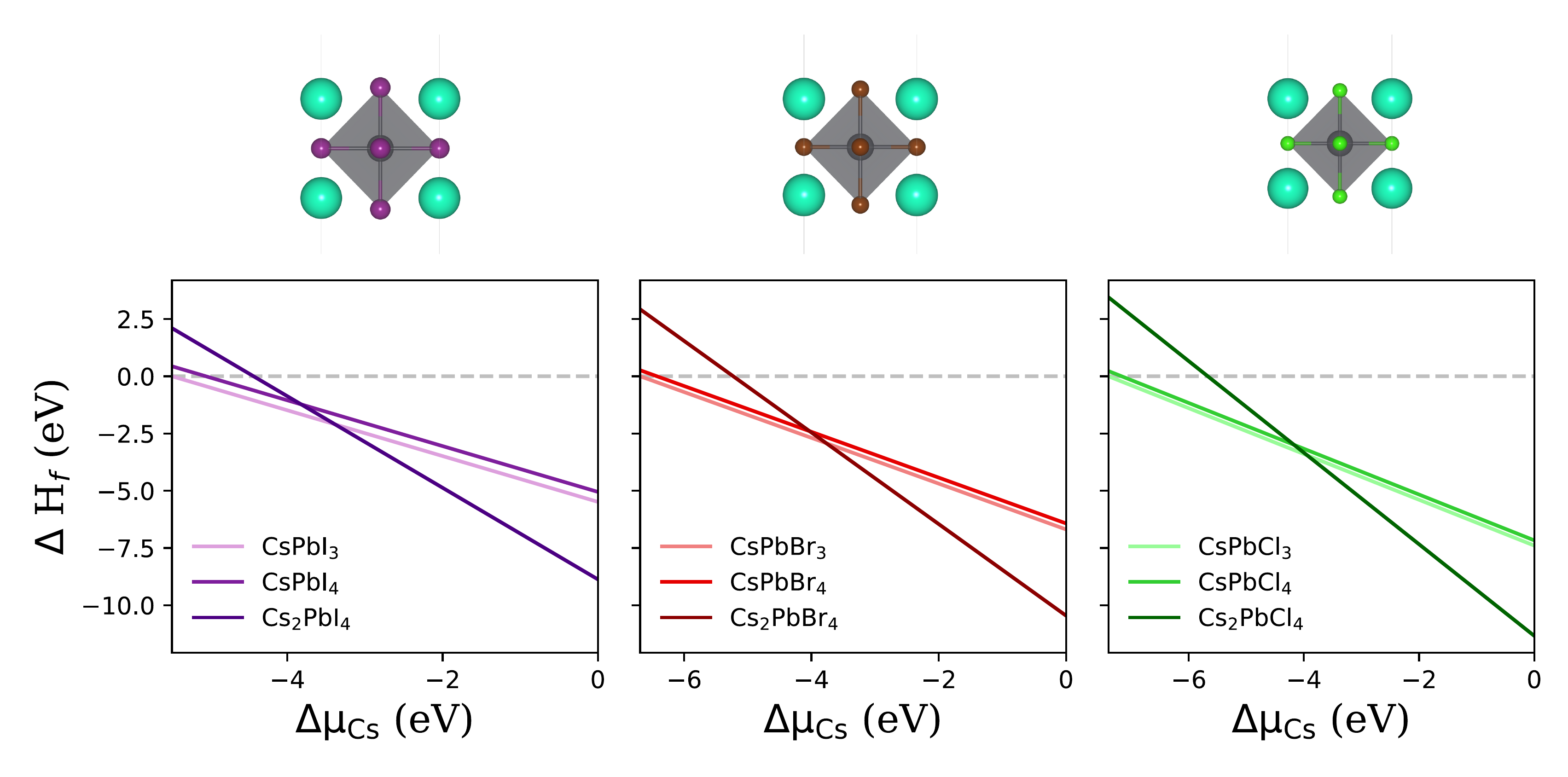}
    \caption{Formation enthalpy depending on the chemical potential of cesium for the CsPbX$_3$, CsPbX$_4$, and Cs$_2$PbX$_4$ monolayers.}
    \label{fig:thermo5}
\end{figure}

\begin{figure}[h]
    \centering
    \includegraphics[width=1\linewidth]{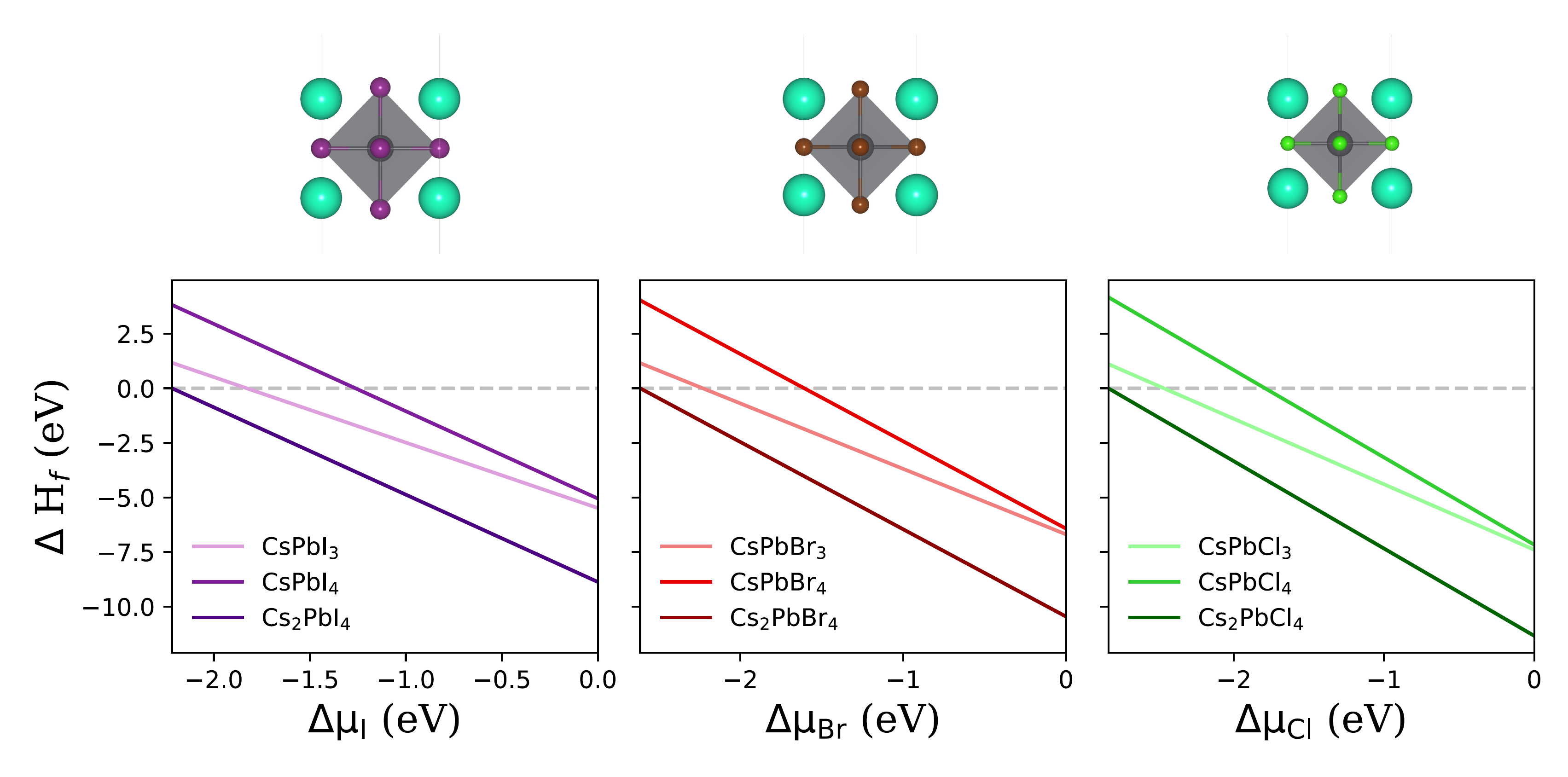}
    \caption{Formation enthalpy depending on the chemical potential of halide for the CsPbX$_3$, CsPbX$_4$, and Cs$_2$PbX$_4$ monolayers.}
    \label{fig:thermo6}
\end{figure}

This section presents the thermodynamical stability analysis of halide perovskite monolayers and their polymorphic phases. Figures~\ref{fig:thermo1}--\ref{fig:thermo6} summarize the decomposition energies, phase competition, and energetic trends as a function of composition and structural symmetry, providing a comprehensive picture of the relative stability of the investigated perovskenes.

\clearpage

\section{Mechanical properties}

The mechanical response of halide perovskite monolayers was evaluated through their elastic tensors. Direction-dependent Young’s modulus and Poisson’s ratio were obtained using Eqs.~(\ref{young_eq}) and~(\ref{poisson eq}), allowing the assessment of elastic anisotropy and mechanical stability across different compositions and polymorphic phases. The table \ref{tab:mechprops} shown the mechanical properties for all of the monolayers.

\begin{equation}
\frac{1}{Y(\theta)} = s_{11} \cos^4\theta + s_{22} \sin^4\theta + 2 s_{13} \cos^3\theta \sin \theta + 2 s_{23} \cos \theta \sin^3\theta + \left(2 s_{12} + s_{33}\right) \sin^2\theta \cos^2\theta
\label{young_eq}
\end{equation}

\begin{equation}
\frac{\mu(\theta)}{Y(\theta)} = (s_{33} - s_{11} - s_{22}) \sin^2\theta \cos^2\theta - s_{12} (\cos^4\theta + \sin^4\theta) + (s_{23} - s_{13})(\cos^3\theta \sin \theta - \cos \theta \sin^3\theta)
\label{poisson eq}
\end{equation}

\begin{table}
    \centering
    \setlength{\tabcolsep}{1pt}
    \caption{Elastic properties of halide perovskite monolayers. The Young's modulus (Y$_{2D}$), Shear modulus (G$^{2D}$), and Layer modulus (Lm) are in $N \cdot \mathrm{m}^{-1}$. The Poisson’s ratio ($\nu$) is dimensionless. The last two rows indicate if the monolayers are mechanically stable, according to the elastic stability conditions for 2D materials. The MS- indicates the M-Square phase, PR- indicates the P-Rectangular phase, PS- corresponds to the P-Square phase, and PO- indicates the P-Oblique phase.}
    \begin{tabular}{cccccccccc}
    \hline
    Property & CsPbI$_3$ & CsPbI$_4$ & Cs$_2$PbI$_4$ & CsPbBr$_3$ & CsPbBr$_4$ & Cs$_2$PbBr$_4$ & CsPbCl$_3$ & CsPbCl$_4$ & Cs$_2$PbCl$_4$ \\
    \hline
    MS-Y$^{2D}$ & 23.96 & 12.83 & 33.76 & 28.32 & 21.72 & 37.98 & 29.25 & 26.50 & 42.82 \\
    PR-Y$^{2D}$ & 8.07 & 3.39 & 8.45 & 8.86 & 9.03 & 9.43 & 9.62 & 9.80 & 10.66 \\
    PS-Y$^{2D}$ & 22.85 & 19.62 & 19.37 & 23.95 & 3.99 & 22.80 & 19.97 & -19.43 & 22.72 \\
    PO-Y$^{2D}$ & 19.98 & -20.73 & 22.56 & 22.93 & 3.06 & 26.18 & 25.50 & 21.72 & 12.85 \\
    MS-$\nu^{2D}$ & -0.05 & -0.14 & -0.00 & 0.01 & 0.04 & 0.01 & 0.02 & 0.09 & 0.04 \\
    PR-$\nu^{2D}$ & 0.67 & -0.29 & 0.69 & 0.69 & 0.57 & 0.67 & 0.67 & 0.67 & 0.62 \\
    PS-$\nu^{2D}$ & 0.05 & 2.75 & -0.03 & 0.05 & 1.64 & 0.08 & 0.28 & -0.16 & -0.07 \\
    PO-$\nu^{2D}$ & 0.15 & 1.15 & 0.16 & 0.15 & 0.96 & 0.12 & 0.15 & 0.05 & -0.33 \\
    MS-Lm & 11.45 & 5.63 & 16.81 & 14.30 & 11.27 & 19.17 & 14.87 & 14.63 & 22.28 \\
    PR-Lm & 12.20 & 1.41 & 12.31 & 13.99 & 10.60 & 13.18 & 14.54 & 14.80 & 13.32 \\
    PS-Lm & 12.01 & -12.38 & 9.44 & 12.65 & -11.66 & 12.33 & 13.74 & -11.49 & 10.58 \\
    PO-Lm & 11.72 & 2.80 & 13.47 & 13.48 & 11.30 & 14.90 & 15.04 & 11.43 & 9.67 \\
    \hline
    \end{tabular}
    \label{tab:mechprops}
\end{table}

\clearpage

\section{Electronic properties}

\begin{figure}[h]
    \centering
    \includegraphics[width=1\linewidth]{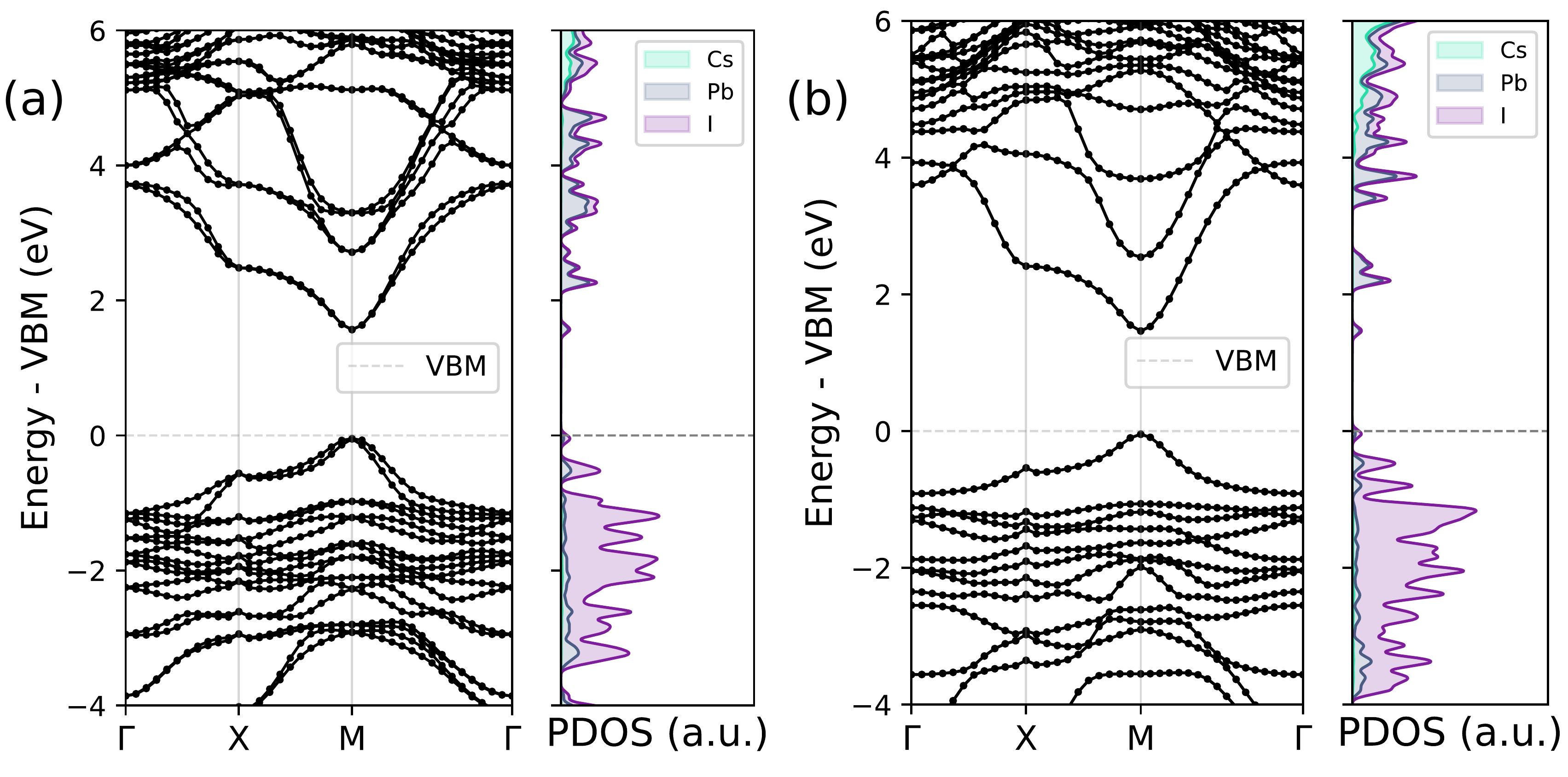}
    \caption{Electronic band structure of the monomorphic CsPbI$_3$ and Cs$_2$PbI$_4$ monolayers calculated including spin--orbit coupling.}
    \label{fig:elec1}
\end{figure}

\begin{figure}[h]
    \centering
    \includegraphics[width=1\linewidth]{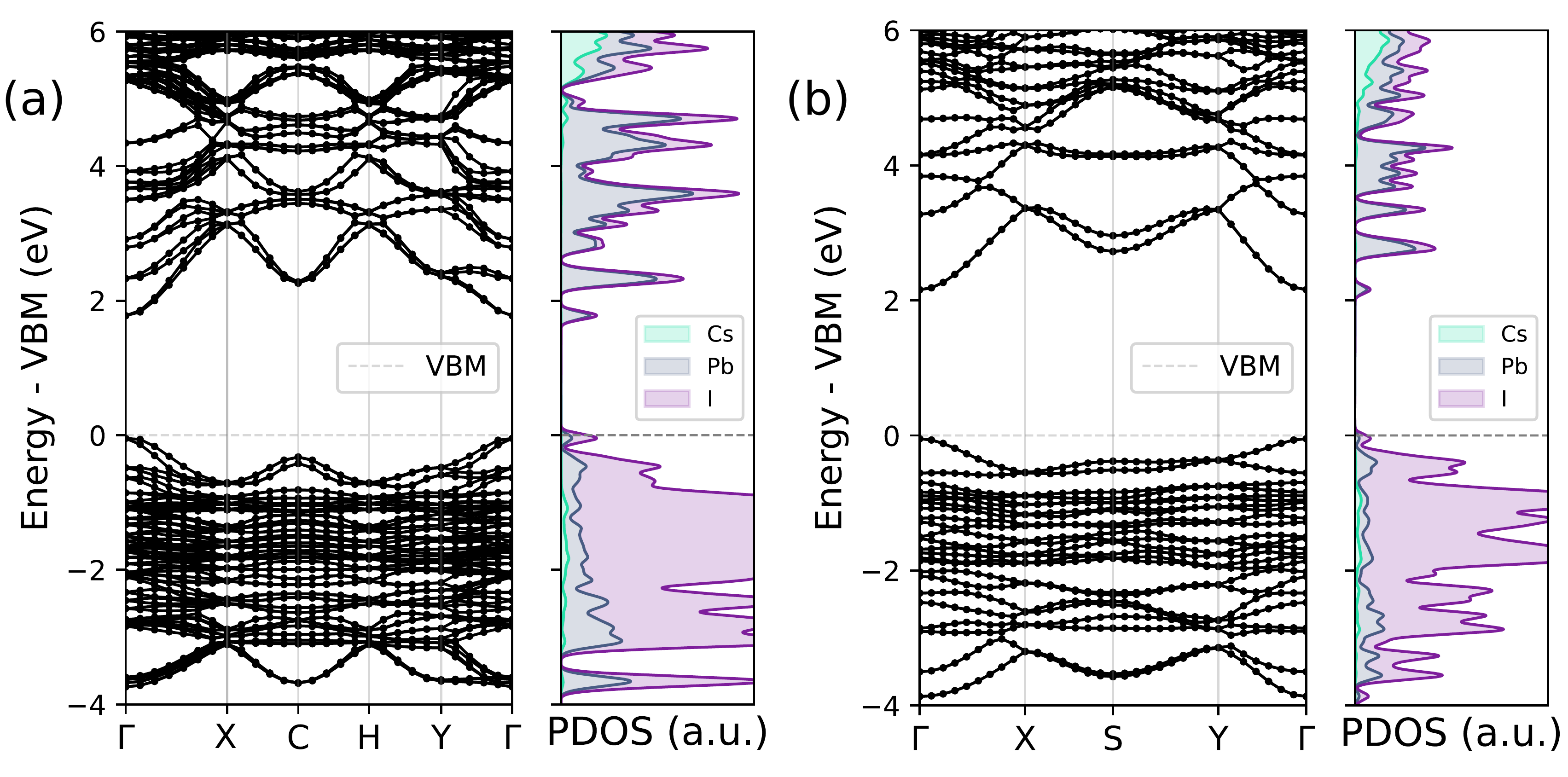}
    \caption{Electronic band structure of the stablest polymorphic phases of CsPbI$_3$ (P-Oblique) and Cs$_2$PbI$_4$ (P-Rectangular), illustrating the impact of structural symmetry on band dispersion and band gap.}
    \label{fig:elec2}
\end{figure}

\begin{figure}[h]
    \centering
    \includegraphics[width=1\linewidth]{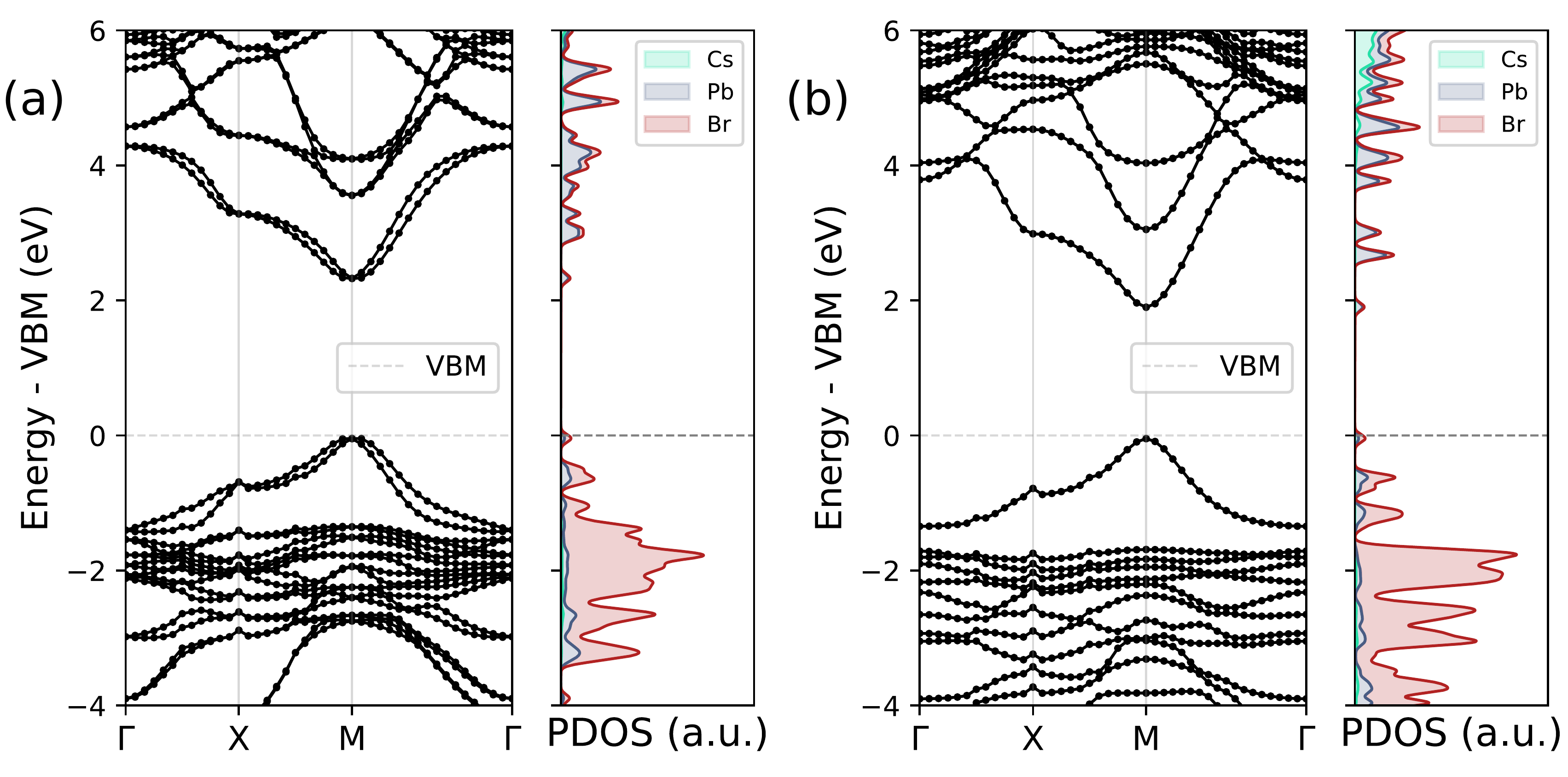}
    \caption{Electronic band structure of monomorphic CsPbBr$_3$ and Cs$_2$PbBr$_4$ monolayers, highlighting the increase in band gap relative to iodide-based systems.}
    \label{fig:elec3}
\end{figure}

\begin{figure}[h]
    \centering
    \includegraphics[width=1\linewidth]{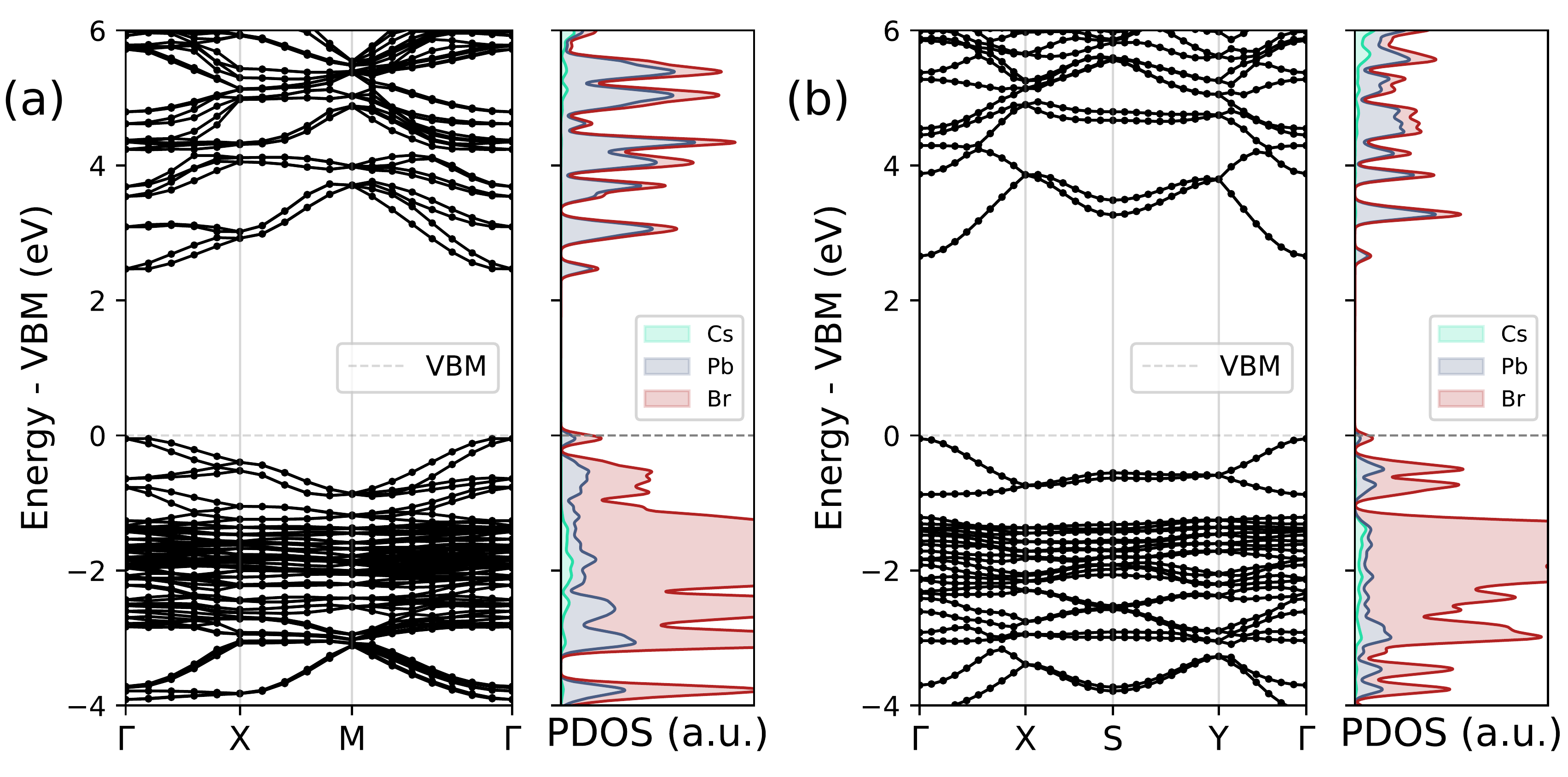}
    \caption{Electronic band structure of the stablest polymorphic CsPbBr$_3$ (P-Oblique) and Cs$_2$PbBr$_4$ (P-Rectangular) monolayers for the most stable non-square phases.}
    \label{fig:elec4}
\end{figure}

\begin{figure}[h]
    \centering
    \includegraphics[width=1\linewidth]{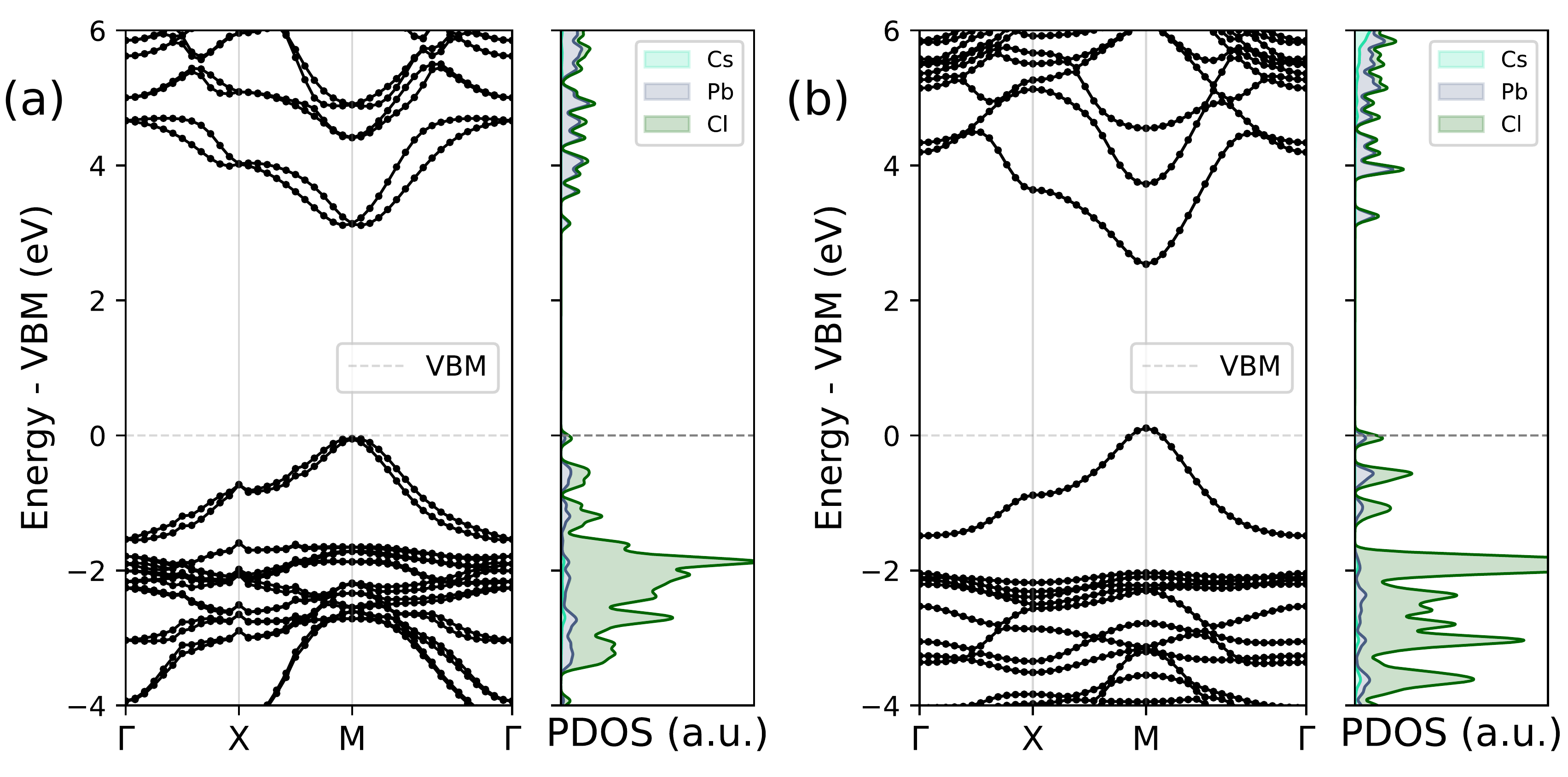}
    \caption{Electronic band structure of monomorphic CsPbCl$_3$ and Cs$_2$PbCl$_4$ monolayers, showing wide band gaps characteristic of chloride-based perovskenes.}
    \label{fig:elec5}
\end{figure}

\begin{figure}[h]
    \centering
    \includegraphics[width=1\linewidth]{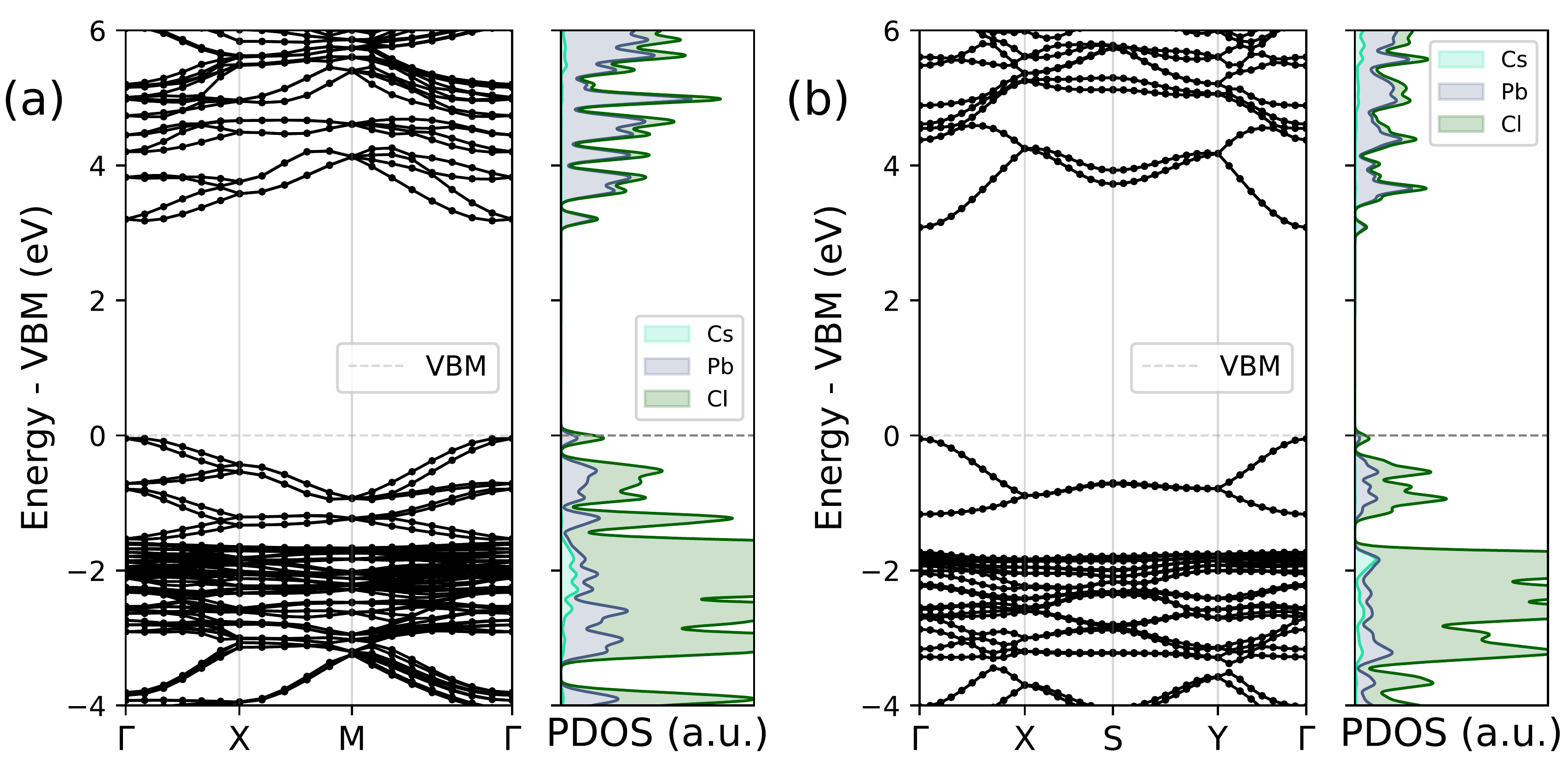}
    \caption{Electronic band structure of the stablest polymorphic CsPbCl$_3$ (P-Oblique) and Cs$_2$PbCl$_4$ (P-Rectangular) monolayers, evidencing minor symmetry-induced modifications near the band edges.}
    \label{fig:elec6}
\end{figure}

This section reports the electronic structure and surface properties of halide perovskite monolayers. Table~\ref{tab:elecprops} summarizes the band gaps and work functions, while Figures~\ref{fig:elec1}--\ref{fig:elec6} present the electronic band structures for the most stable monomorphic and polymorphic phases, highlighting the effects of composition and symmetry on the electronic behavior.

\begin{table}
    \centering 
    \caption{Electronic properties of halide perovskite monolayers. The band gap energy ($\Delta$E$_{g}$), and the work functions ($\Phi_{top}$, and $\Phi_{bottom}$) are in eV. The MS- indicates the M-Square phase, PR- indicates the P-Rectangular phase, PS- corresponds to the P-Square phase, and PO- indicates the P-Oblique phase.} 
    \begin{tabular}{lcccccc}
    \hline
     & CsPbI$_3$ & Cs$_2$PbI$_4$ & CsPbBr$_3$ & Cs$_2$PbBr$_4$ & CsPbCl$_3$ & Cs$_2$PbCl$_4$ \\
    \hline
    MS-$\Delta$E$_{g}$ & 1.61 & 1.51 & 2.37 & 1.95 & 3.16 & 2.43 \\
    PR-$\Delta$E$_{g}$ & 1.79 & 2.21 & 2.50 & 2.70 & 3.22 & 3.13 \\
    PS-$\Delta$E$_{g}$ & 1.78 & 1.74 & 2.46 & 2.19 & 3.19 & 2.64 \\
    PO-$\Delta$E$_{g}$ & 1.82 & 1.77 & 2.51 & 2.21 & 3.22 & 2.63 \\
    MS-$\Phi_{top}$ & 5.06 & 5.01 & 5.35 & 4.96 & 5.53 & 5.35 \\
    PR-$\Phi_{top}$ & 5.44 & 5.04 & 5.65 & 5.14 & 5.79 & 5.10 \\
    PS-$\Phi_{top}$ & 5.23 & 5.32 & 5.47 & 5.28 & 5.62 & 5.15 \\
    PO-$\Phi_{top}$ & 5.26 & 5.30 & 5.50 & 5.28 & 5.65 & 5.19 \\
    MS-$\Phi_{bottom}$ & 5.75 & 5.01 & 5.95 & 4.97 & 6.04 & 5.35 \\
    PR-$\Phi_{bottom}$ & 5.58 & 5.08 & 5.79 & 5.18 & 5.90 & 5.15 \\
    PS-$\Phi_{bottom}$ & 5.87 & 5.33 & 6.06 & 5.30 & 6.15 & 5.18 \\
    PO-$\Phi_{bottom}$ & 5.89 & 5.31 & 6.09 & 5.29 & 6.18 & 5.22 \\
    \hline
    \end{tabular}
    \label{tab:elecprops} 
\end{table}

\clearpage
